\documentclass[%
reprint,
amsmath,amssymb,
aps,
prd,
]{revtex4-1}

\usepackage{xcolor}
\usepackage[capitalize]{cleveref}
\usepackage{graphicx}
\usepackage{dcolumn}
\usepackage{bm}
\usepackage{tabularx}
\usepackage{ragged2e}
\usepackage{subcaption}
\usepackage{acronym}
\newcommand\mycaption[2]{\caption{\textbf{#1}\newline\footnotesize#2}}

\newcommand\kagra{KAGRA}
\newcommand\Pmin{\mathrm{min} }
\newcommand\Pmax{\mathrm{max} }
\newcommand\Pbest{\mathrm{best} }
\newcommand\Bilby{\texttt{Bilby} }
\newcommand\dynesty{\texttt{dynesty} }
\newcommand\lal{\texttt{LALSimulation} }

\newacro{BBH}{Binary Black Hole}
\newacro{BNS}{Binary Neutron Star}

\begin{document}

\preprint{APS/123-QED}

\title{Comparing advanced-era interferometric gravitational-wave detector network configurations:
sky localization and source properties}

\author{Mattia Emma}
\email{mattia.emma@ligo.org}
\author{Tiago Fernandes de Nobrega } 
\author{Gregory Ashton}
\affiliation{Department of Physics, Royal Holloway, Egham, TW20 0EX}

\date{\today}

\begin{abstract}
The expansion and upgrade of the global network of ground-based gravitational wave
detectors promises to improve our capacity to infer the sky-localization of transient sources,
enabling more effective multi-messenger follow-ups. At the same time, 
the increase in the signal-to-noise ratio of detected events allows for more precise estimates
of the source parameters. This study aims to assess the performance of advanced-era networks of 
ground-based detectors, focusing on the Hanford, Livingston, Virgo, and KAGRA instruments. 
We use full Bayesian parameter estimation procedures to predict the scientific potential of a network. 
Assuming a fixed LIGO configuration, we find that the addition of the Virgo detector is beneficial 
to the sky localization starting from a binary neutron star horizon distance of 20~Mpc and improves
significantly from 40~Mpc onwards for both a single and double LIGO detector network, 
reducing the inferred mean sky-area by up to 95\%. Similarly, the KAGRA detector tightens the constraints,
starting from a sensitivity range of 10~Mpc. Looking at highly-spinning binary black holes,
we find significant improvements with increasing sensitivity in constraining the intrinsic 
source parameters when adding Virgo to the two LIGO detectors. 
Finally, we also examine the impact of the low-frequency cut-off data on the signal-to-noise ratio. 
We find that existing 20~Hz thresholds are sufficient and propose a metric to monitor this to study detector performance. Our findings quantify how future enhancements in detector sensitivity 
and network configurations will improve the localization of gravitational wave sources and allow for more 
precise identification of their intrinsic properties.
\end{abstract}

\maketitle

\section{\label{sec:Intro} Introduction}
The detections of the first \ac{BBH}, GW150914~\cite{LIGOScientific:2016aoc}, 
and \ac{BNS}, GW170817~\cite{LIGOScientific:2017vwq}, 
gravitational wave signals have been major discoveries opening a new window into our Universe. 
They not only allowed for the confirmation of Einstein's theory of general relativity at an unprecedented 
level of accuracy~\citep{LIGOScientific:2019fpa, LIGOScientific:2020tif, LIGOScientific:2021sio}, 
but also deepened our knowledge of the physics of compact 
objects~\citep{LIGOScientific:2018hze,Margalit:2017dij, LIGOScientific:2018cki, Nair:2019iur} 
and the evolutionary history of the Universe~\citep{LIGOScientific:2017adf}. 
These scientific breakthroughs were only possible thanks to the highly sophisticated 
network of ground-based gravitational wave interferometers developed by the LIGO 
Scientific~\citep{LIGOScientific:2014pky}, Virgo~\citep{VIRGO:2014yos} and \kagra~\citep{Kagra:2020tym} 
collaborations. The progressive development of the detectors in the past decades has allowed for a 
continuous increase in the detection rate throughout the first three observing 
runs~\citep{LIGOScientific:2018mvr, LIGOScientific:2020ibl, LIGOScientific:2021djp}. 
This trend is expected to continue in the current fourth observing run (O4) and beyond, 
as detector upgrades and the development of next-generation 
instruments~\cite{Unnikrishnan:2013qwa, Reitze:2019iox, Punturo:2010zz} promise further 
advancements in sensitivity and precision.

When considering networks of ground-based gravitational wave interferometers, 
one of the metrics used to quantify their performance for transient sources is the 
sky localization accuracy obtained through triangulation~\cite{Schutz:2011tw}. 
Constraining the position of the source of a detected signal improves our understanding of the 
different populations of compact objects, shedding light into their distribution across the 
Universe~\citep{LIGOScientific:2018jsj, LIGOScientific:2020kqk, Kagra:2021duu}. 
It also improves the multimessenger follow-up capability~\cite{LIGOScientific:2016qac, AghaeiAbchouyeh:2023lap}. 
The coincident detection of GW170817, GRB170817A and AT2017gfo~\cite{LIGOScientific:2017ync, LIGOScientific:2017zic}
showed the incredible potential of multimessenger detections, allowing for determining the origin of Gamma Ray 
Bursts, producing an independent measurement of the Hubble 
constant~\cite{LIGOScientific:2017adf, Dietrich:2020efo, Bulla:2022ppy} and enhancing our knowledge of the 
synthesis of heavy elements~\cite{Kasen:2017sxr, Drout:2017ijr}. GW170817 also proved the potential of multi-detector detections. The additional Virgo data, even if the signal was beyond 
its binary neutron star horizon allowed for a significant refinement of the sky-localization area and the source 
parameter estimates~\citep{LIGOScientific:2017vwq}. Future detections of multimessenger counterparts could 
greatly improve our constraints on the neutron star equation of state~\citep{LIGOScientific:2018cki} 
(e.g.,~\citet{Koehn:2024set} for a review)
and our understanding of the formation and evolution of binary black 
holes~\citep{Bartos:2016dgn, Graham:2020gwr, Tagawa:2023uqa, Zhou:2023rtr}. 
To evaluate the electromagnetic follow-up capability, it is important to consider 
that the field of view of current optical telescopes ranges from 35~deg$^2$, with an R band sensitivity 
of $\sim$20~mag, for the Zwicky Transient Facility (ZTF)~\citep{Masci:2018neq, Graham:2019qsw} to the 9.6~deg$^2$ 
at $\sim$24.5~mag for the Vera Rubin Telescope~\citep{LSST:2008ijt,LSSTScience:2009jmu,Nissanke:2012dj}.

In this work, we quantify the benefits of having a three- or four-ground-based detector network, 
including \kagra~and Virgo, compared to the sole LIGO detectors, 
focusing on the refinements in source-localization accuracy. Starting with~\citet{Jaranowski:1996hs}, 
several authors have worked on the estimation of sky-localization, and more generally parameter estimation 
accuracy for different networks of detectors. \citet{Schutz:2011tw}, introduced three figures of merit to 
compare the performance of networks of detectors. \citet{Fairhurst:2009tc, Fairhurst:2010is} then focused on 
analytically computing the improvements in sky-localization constraints using triangulation from the timing information. \citet{Berry:2014jja} then looked at the results obtained with this method for different networks of detectors and compared them to parameter estimation ones.
\citet{Nissanke:2012dj} first employed Bayesian methods to compare networks of second-generation ground-based 
detectors including a possible detector in Australia~\citep{Bailes:2019oma}.
Successive works have either focused on using a limited amount of information to determine 
the sky-localization~\cite{Wen:2010cr} or on quantifying improvements for specific networks once 
the component detectors have reached 
design-sensitivity~\cite{Klimenko:2011hz, Veitch:2012df, Kagra:2013rdx, Pankow:2018phc, Pankow:2019oxl, Shukla:2023kuj}. \citet{Singer:2014qca} first looked at the impact of the Virgo detector on the HL network.
Furthermore, some studies have concentrated on evaluating the best locations for the third-generation 
ground-based detectors~\citep{Weiss:2010, Raffai:2013yt, Zhao:2017cbb, Hall:2019xmm}. 
Here, we specifically investigate the enhancements facilitated by the inclusion of the Virgo and \kagra~detectors at the sensitivity levels projected for O4 and the fifth observing run, 
O5~\citep{Kagra:2022qtq,LIGO-ObserverDocs}. 

However, sky-localization is not the only improvement one obtains from a better network.
We also expect to see improvements in measurements of the source parameters. 
Most of the signals detected to date originate from black hole binaries with low spin magnitudes 
($\chi_i\lesssim 0.4$), aligned spins, and comparable mass-components ($q\geq0.5$)~\citep{Kagra:2021duu}. 
These results would favor theories that suggest an isolated evolution 
scenario~\citep{Dominik:2014yma, Neijssel:2019irh, Marchant:2021hiv} as the main formation channel 
of black hole binaries~\citep{Zevin:2020gbd}. According to this model even if after their formation 
the black holes have misaligned spins, their unhindered evolution and interaction would lead to a 
progressive alignment of the spins on time scales much shorter than the merger time. 
The support for precession, high mass ratios, and high spin magnitudes in the analysis of events such as 
GW190412~\citep{LIGOScientific:2020stg}, GW190814~\citep{LIGOScientific:2020zkf}, 
GW190521~\citep{LIGOScientific:2020iuh, LIGOScientific:2020ufj}, and GW200129\_065458~\citep{KAGRA:2021vkt, Hannam:2021pit}, hereafter referred to as GW200129, 
has challenged the models for the formation of compact-object binaries~\citep{Olejak:2020oel} and 
hinted at a connection between precessing systems and high mass ratios. Alternative formation channels 
for these systems include dynamical formation~\citep{Fragione:2018yrb, Tagawa:2019osr}, in environments 
with high stellar density, hierarchical mergers~\citep{Fragione:2019hqt, Vigna-Gomez:2020fvw} and chemically 
homogeneous evolution~\citep{Mandel:2015qlu, deMink:2016vkw, duBuisson:2020asn}. Unfortunately, 
the accuracy in the determination of precession and the intrinsic source parameters, 
e.g., masses, spins, and their combinations have been limited by the difficulties in producing 
precise waveform-approximants in the region of the multi-dimensional parameter space where high-spins 
and high mass ratios intersect~\citep{Colleoni:2020tgc, Pratten:2020igi, Hannam:2021pit, Ramos-Buades:2023ehm, Thompson:2023ase, Dhani:2024jja}, and by the signal-to-noise 
ratio with which we detect the signals~\citep{Vecchio:2003tn, Lang:2006bsg, Vitale:2014mka, Green:2020ptm}.

In this work, we vary the sensitivity of the Virgo and \kagra~detectors and compare how these add to the 
network performance, assuming a fixed LIGO sensitivity.
Throughout this study, we refer to the different networks of detectors using abbreviations derived from 
the initial letters of the included interferometers' names. For instance, HL refers to the Hanford and 
Livingston detectors, following the convention from~\citet{Kagra:2013rdx}.
In Section~\ref{sec:Method}, we detail the simulation methodology and the post-processing procedure to 
perform full parameter estimation and infer the relevant parameters for our analysis. In Section~\ref{sec:bbh} 
and~\ref{sec:bns}, we present the results of the evolution of the sky localization area while varying the 
sensitivity range of one of the detectors in the specific network for binary black hole and binary neutron star 
gravitational wave signals, respectively. We limit our study to the LK, LV, HLK, HLV, and HLVK detector networks.
We include the two detector networks to account for the variable duty factor of the single interferometers, 
which could result in only a sub-group of detectors of the available network being on duty at the time an event 
happens~\citep{LIGO:2021ppb, Nitz:2020naa}. In Section~\ref{sec:spinning}, we look at the changes in the 
parameter estimation results for a high spin binary black holes merger, emulating the spin magnitude values of 
GW200129~\cite{Hannam:2021pit}.
Finally, we study the minimum-frequency cut-off in Section~\ref{sec:min_freq}. The signal-to-noise ratio of 
detected events, which determines the accuracy of source-parameter estimates, is limited not only by the 
sensitivity of the detectors, but also by the duration of the signal falling inside the detector's sensitivity 
bandwidth. The lower cut-off frequency for the data used in the parameter estimation of events detected during 
the third observing run was typically 20 Hz~\cite{LIGOScientific:2019hgc, LIGOScientific:2021djp}, a decision 
made to balance gains against exponentially increased analysis times.
In Section~\ref{sec:min_freq}, we investigate this choice of minimum frequency and develop a metric 
to quantify the loss of signal-to-noise ratio. We then apply this to data from the third observing run. 
Finally, in Section~\ref{sec:discussion} and~\ref{sec:conclusion}, we discuss and summarize our results.

\section{\label{sec:Method} Simulation setup}

\begin{figure}[htp!]
\includegraphics[width=0.45\textwidth]{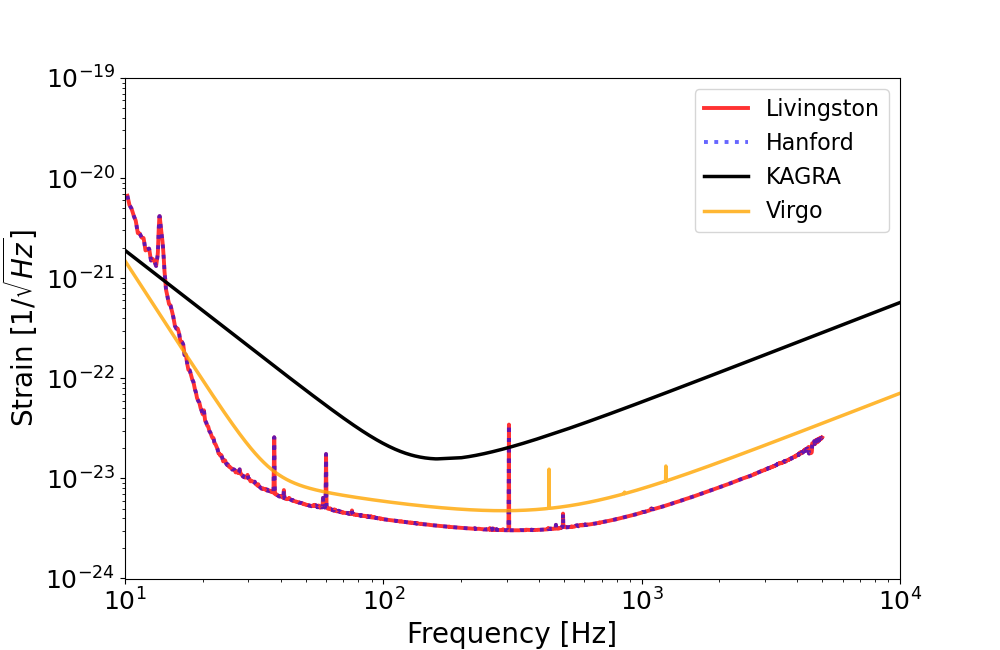}
\caption{ \RaggedRight \justifying Comparison of representative sensitivity curves of the 
detectors employed in our simulations. The overlapping blue dotted and red full lines are the sensitivity 
curves of the Hanford and Livingston detectors with an optimal binary neutron star sensitivity range of 180~Mpc. 
The \kagra~curve, in black, and the Virgo one, in yellow, correspond to binary neutron star detection ranges 
of 25 and 115~Mpc, respectively. An equal mass neutron star binary with component masses of 1.4~M$_\odot$ was 
used to compute the horizon distance~\cite{Chen:2017wpg}.}   
\label{fig:asd}
\end{figure}

Our objective is to simulate the parameter estimation of gravitational wave signals 
by fully replicating the process involved in a real event excluding the impact of glitches, 
i.e., non-Gaussian transient noise. 
We add simulated signals to simulated colored Gaussian noise generated from a PSD. 
Two methods are generally employed to obtain PSD curves in the literature. 
The first method, utilizes actual detector data, as done with the data from the first three 
observing runs~\cite{Kagra:2013rdx}. The second method consists of computing and combining analytical 
noise curves for different sources influencing each considered detector~\cite{VIRGO:2014yos}. 
Notably, the latter method allows for the simulation of PSDs for future detectors and upgraded versions of 
current detectors lacking empirical data. These simulations effectively incorporate broken power laws 
with additional lines, such as those corresponding to the frequency of the power grid coupled to the detector 
(e.g., 60~Hz in the US) and other known technical noise sources.
In this study, we utilize publicly available simulated sensitivity curves for O4 LIGO, \kagra, and Virgo, 
as presented in~\citet{Kagra:2013rdx}. Specifically, we employ the LIGO O4 high-sensitivity curve with a 
horizon distance of 180~Mpc, the Virgo high-sensitivity curve utilized for O4 simulations with a horizon 
distance of 115~Mpc, and the 25~Mpc \kagra~sensitivity curve, shown in Fig.~\ref{fig:asd}.

To investigate the general trend of sky localization with respect to sensitivity range,
we scale the sensitivity curves to the desired range. To do so, we multiply them by an arbitrary 
calibration factor until achieving that range with an error lower than 10$^{-2}$. 
We determine the horizon distance of the sensitivity curve using methods developed 
by~\citet{Chen:2017wpg}, simulating an equal-mass binary neutron star with component masses of 
1.4~M$_\odot$. For \kagra, we consider sensitivity range values between 5 and 25~Mpc, while for Virgo, 
we employ sensitivity curves ranging from 10 to 180~Mpc, approximately covering the expected 
sensitivity ranges of the two detectors for future observation 
runs~\cite{Kagra:2013rdx,Kagra:2022qtq,Kagra:2022fgc, VIRGO:2023elp}. 
Although the scaling method employed for the spectral density curves provides only 
an approximation of the real curves, it allows for a first estimate of the performance of the detector 
networks, which could be refined in the future either with detector curves from real data for the specific horizon distance values or using a different scaling for each power law composing the PSD, e.g., a bigger scaling factor at lower frequencies for which the most improvements are expected soon. 
 
\begin{figure*}[t]
\includegraphics[width=0.8\textwidth]{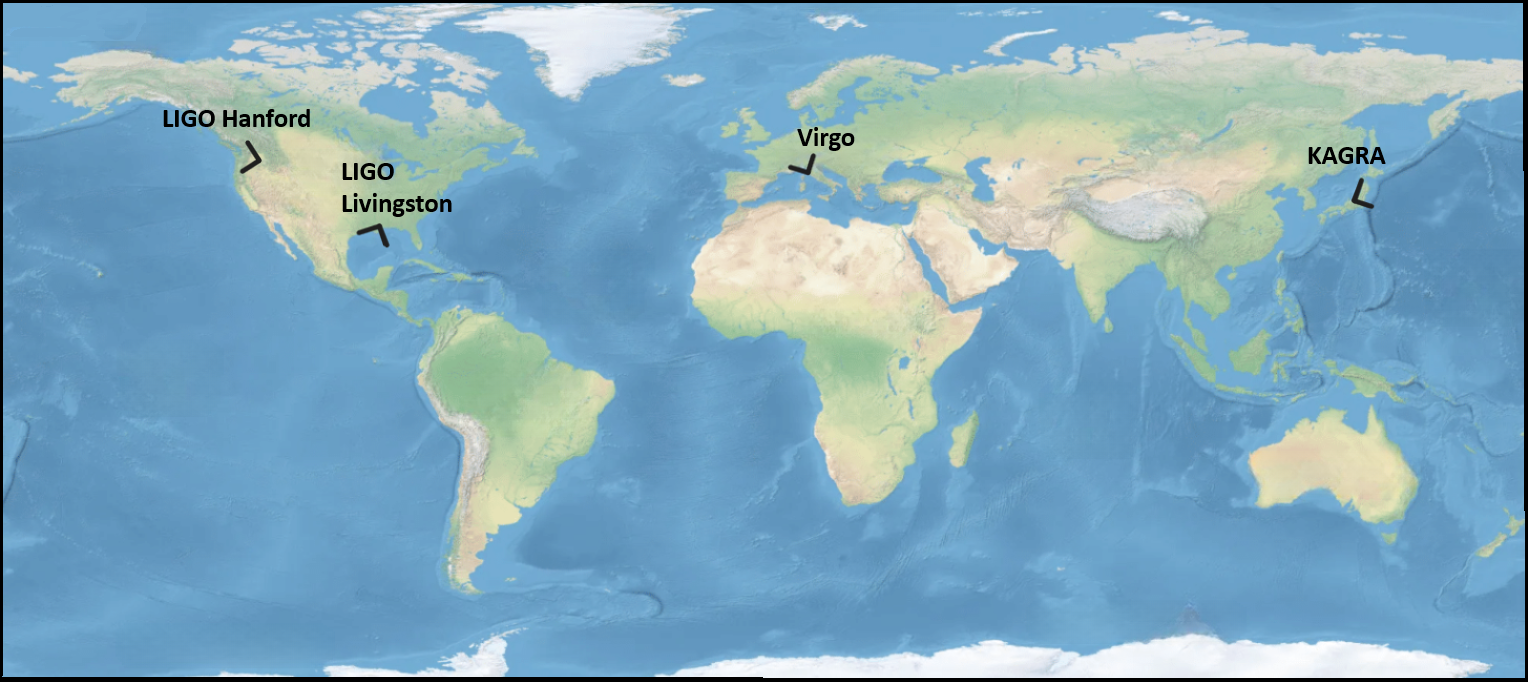}    
\caption{\RaggedRight \justifying Location and orientation of the current network of gravitational-wave 
observatories, identified by the solid black lines. We omit GEO600 as we do not include it in our analysis.}
\label{fig:det_orientation}
\end{figure*}

Our primary focus is on assessing the impact of the \kagra~and Virgo detectors on the parameter 
estimation of gravitational wave signals assuming a fixed O4 LIGO network. 
To this end, we define the simulated source's location relative to these two additional 
interferometers and vary only their sensitivity range while keeping that of the LIGO detectors 
constant at 180~Mpc.  We use the antenna power pattern function $P(\theta,\phi)$ defined in~\citet{Schutz:2011tw} as:
\begin{align}
    P(\theta,\phi)&=F_+(\theta,\phi,\psi)^2+F_{\times}(\theta,\phi,\psi)^2
    \label{eq:antenna_response}\\
    &= \frac{1}{4}(1+\cos^2\theta)^2\cos^22\phi+\cos^2\theta\sin^22\psi,                
\label{eq:antenna_power}
\end{align}
where $F_+$ and $F_\times$ are called the sensitivity functions and $\theta$ and $\psi$ are the spherical 
coordinates with respect to the detector's axes. We note that the antenna power pattern does not depend on 
the angle $\psi$ related to the polarization of the gravitational wave. 
It depends only on the relative orientation of the detector to the source, 
so that we obtain a wider sky coverage when interferometers are spread across the globe and have maximally 
different orientations. Table~\ref{tab:localization} presents the right ascension (RA), declination (DEC), and single
detector antenna pattern function values
for the source localizations maximizing and minimizing the antenna power pattern function
for the \kagra~and Virgo detectors. We assume $\psi=0$ in the computation of the antenna power pattern function
values. We show the location and orientation of the current network of ground-based detectors in 
Figure~\ref{fig:det_orientation}. We omit the GEO600~\citep{Willke:2002bs} detector as we do not include
it in our analysis. The large distance 
between Virgo and \kagra~makes these extremely useful for sky-localization purposes in a four-detector network to obtain 
precise time-delay measurements.

\begin{figure*}[t]
\includegraphics[width=0.4\textwidth]{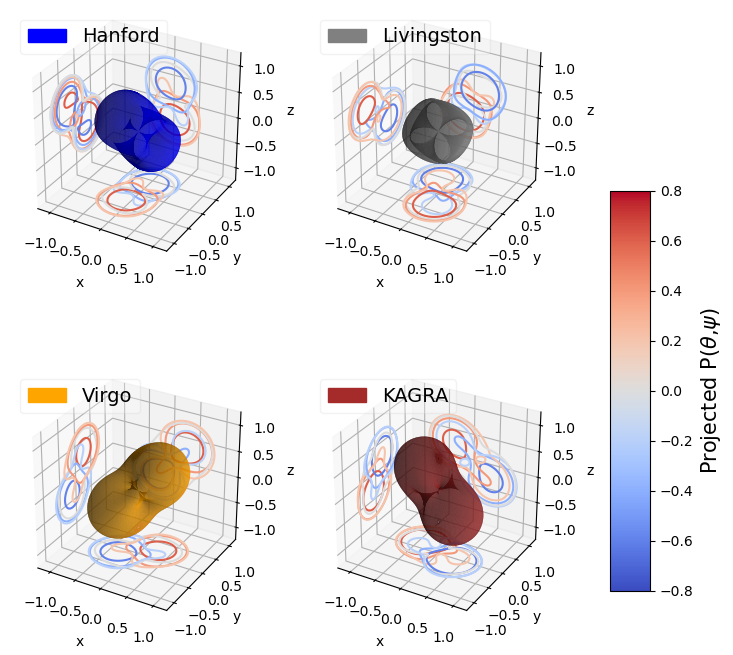}  
\includegraphics[width=0.42\textwidth]{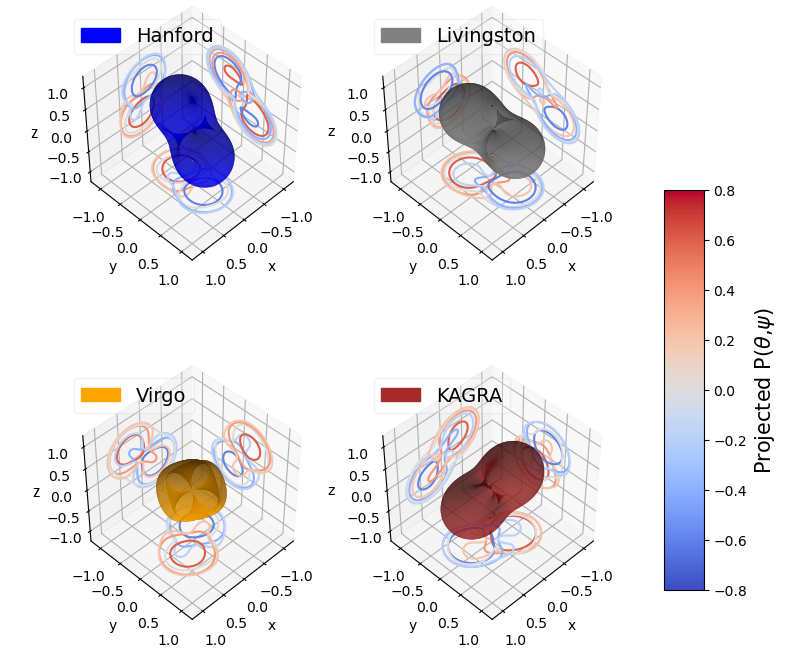}
\includegraphics[width=0.45\textwidth]{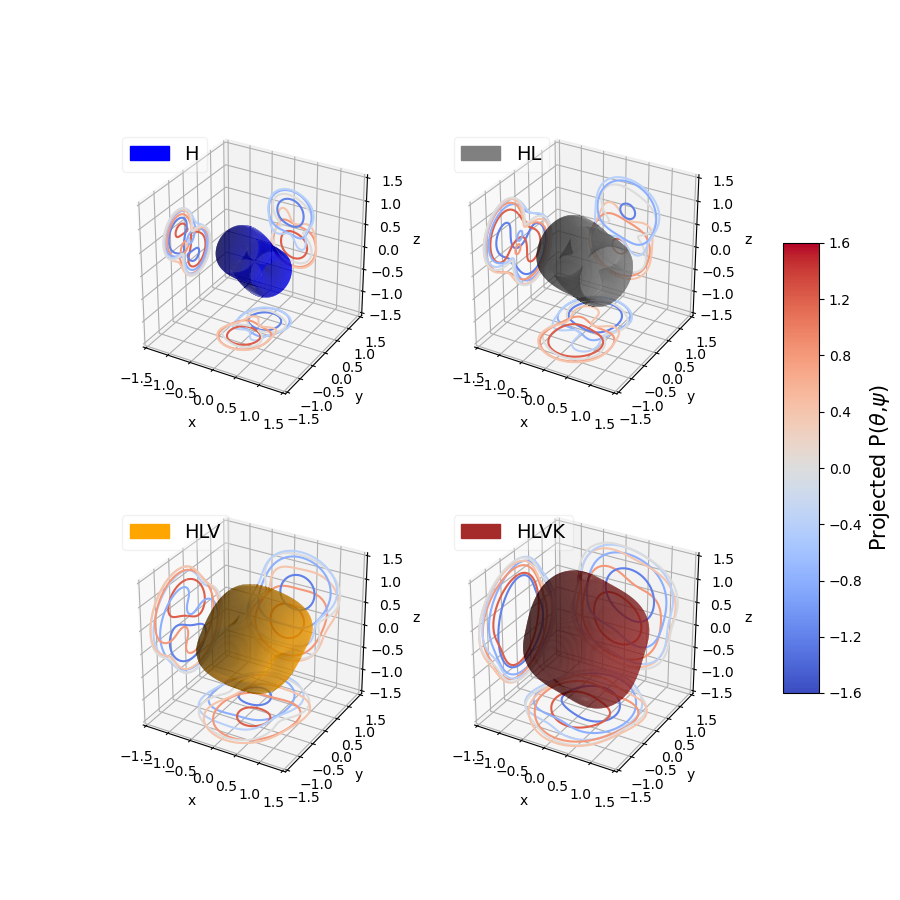}  
\includegraphics[width=0.45\textwidth]{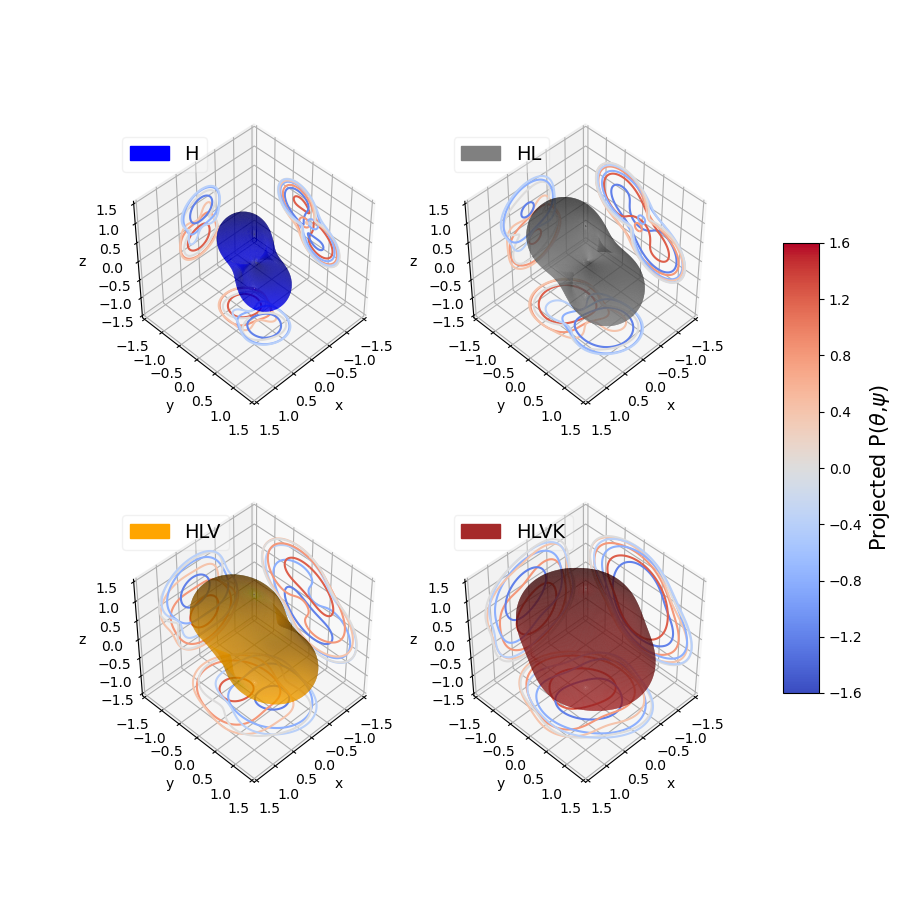}
\caption{\RaggedRight \justifying 3D visualizations of the antenna power pattern amplitude, 
from Eq.~\ref{eq:antenna_power}, for the LIGO-Hanford, LIGO-Livingston, Virgo and \kagra~detectors 
at a fixed arbitrary GPS time, so that the relative orientation of the detectors can be compared. The projections 
on the two-dimensional planes are the contours of the antenna power pattern amplitude for the projected axis.}
\label{fig:antenna_power_pattern}
\end{figure*}

Furthermore, in Figure~\ref{fig:antenna_power_pattern} we show 3D visualizations of the antenna power pattern 
amplitude for the four employed detectors at a fixed GPS time from two different angles.
We notice that all detectors have different relative orientations, agreeing with the different antenna pattern values presented
in Table~\ref{tab:localization}. While the two LIGO detectors are almost parallel,
they are orthogonal to KAGRA and Virgo, which are also mutually orthogonal. In this work, 
we focus on the case in which KAGRA and Virgo have lower sensitivities than the two LIGO detectors. 
Their resulting lower signal-to-noise ratio could be partially compensated for some regions of the sky 
by their different relative orientations. 
From Table~\ref{tab:localization}, we can compute an estimate of the fraction of LIGO's sensitivity 
at which Virgo and KAGRA data would begin to have an effect on the sky localization constraints. 
This corresponds to the quotient between the single LIGO and the Virgo/\kagra~antenna power pattern amplitudes 
computed at the right ascension and declination values that maximize Virgo/\kagra's antenna power pattern. 
For Virgo, we find that the quotient is $\simeq 1/6$ so if the LIGO detectors have a sensitivity of 180~Mpc,
we would expect to see significant improvements in the sky localization when Virgo reaches 
sensitivities $\simeq 30$~Mpc. For \kagra, we find a similar quotient so that we expect improvements from 30-35~Mpc
onwards, values which we do not include in this study.
For all the values included in Table~\ref{tab:localization} we used the \Bilby library~\cite{Ashton:2018jfp} to compute the right ascension and declination for the maximum 
and minimum, respectively, P$_{\Pmax}$ and P$_{\Pmin}$, of the antenna power pattern function at a specific 
arbitrary GPS time for each detector. In Appendix~\ref{sec: best network} we report the same values for P$_{\Pbest}$, the sky position maximizing the combined detector network antenna power pattern function for each configuration.
 \\

\begin{table*}[htp!]
\centering
\begin{tabular}{|c|c|c|c|c|c|c|} \hline 

                & RA (rad)  &  DEC (rad)  & P$_{K}$ & P$_{V}$ & P$_{H}$ & P$_{L}$\\ \hline  
\kagra~P$_{\Pmax}$      & 1.709        &   0.761 & 1.0 & 0.14 & 0.22 & 0.17\\ \hline    
\kagra~P$_{\Pmin}$       & 4.422        &   0.888 & 10$^{-10}$ & 0.46 & 0.71 & 0.80\\ \hline  
Virgo P$_{\Pmax}$       & 5.785        &   0.761  & 0.24 & 1.0 & 0.17 & 0.14\\ \hline  
Virgo P$_{\Pmin}$       & 1.392        &   0.318  & 0.83 & 10$^{-10}$ & 0.17 & 0.39\\ \hline

\end{tabular}
\caption{\RaggedRight \justifying Table of the right ascension (RA), declination (DEC) and the values of the single
detector's antenna pattern amplitude for the source 
localization maximizing and minimizing the antenna power pattern function, P$_{\Pmax}$ and P$_{\Pmin}$ respectively, 
for the \kagra~and Virgo detectors. The GPS time is fixed at 1379969683.0. }
\label{tab:localization}
\end{table*}

Given the source location values, to conduct each analysis we simulate a gravitational waveform and add it to a simulated 
interferometer data strain to capture the detector's response (see Table~\ref{tab:simulation} 
in \cref{sec: app_tables} for the simulated parameter values). The process is facilitated using 
\Bilby~\cite{Ashton:2018jfp, Romero-Shaw:2020owr}. The colored Gaussian noise background data is 
simulated from the PSD based on the scaled sensitivity curves, while the added waveform is generated 
using the \lal library~\cite{lalsuite}: we employ the \texttt{IMRPhenomX} family of phenomenological waveform 
models~\cite{Pratten:2020fqn}. We use the simulated detector response and 
waveform to calculate the gravitational wave transient likelihood~\cite{Whittle:1951}. We then provide 
this likelihood to the \Bilby sampler along with priors on relevant parameters, shown in Table~\ref{tab:prior} 
in the \cref{sec: app_tables}, enabling a comprehensive parameter estimation. We opt for the \dynesty 
sampler~\cite{Speagle:2019ivv}, a nested sampling algorithm~\cite{Skilling:2004pqw} that employs 
differential evolution and adaptive sampling for an effective exploration of the high-dimensional parameter space. 
We summarize the configuration details for the \dynesty sampler in Table~\ref{tab:dinesty} in \cref{sec: app_tables}.

\subsection{Post-processing}
\label{sec:post_proc}

 For the gravitational wave signals from highly-spinning binary black hole mergers, we want to understand if a 
 larger and more sensitive detector network would allow us to better constrain the source's intrinsic 
 parameters (see Sec.~\ref{sec:spinning}), while in the non-spinning binary black hole and binary neutron star 
 cases, we want to quantify the improvements in determining the sky localization. We focus on the sky localization area as we expect it to be influenced the most by adding a new detector to the network~\citep{Fairhurst:2010is}. 
 Indeed, the additional detector does not have a comparable or higher sensitivity range that would enhance the 
 source parameter estimation results. Nevertheless, its distant location to the LIGO detectors allows 
 for a significant gain in time-delay information which can be used for triangulation purposes. To this end, from 
 the parameter estimation results of the non-spinning system's mergers, we generate a sky map, a graphical 
 representation of the probable source locations of a gravitational wave event on the celestial sphere.
 We do this by passing the posterior probability samples of the distance, right ascension, and declination to 
 the \texttt{ligo.skymap} function~\cite{ligo_skymap}. Subsequently, we extract the values corresponding to the 
 90\% probability contours of the sky localization area from the generated sky map. To address the stochastic 
 nature of the sampling process and Gaussian noise, and consequently, the variability in sky localization area 
 results, we conduct multiple simulations with identical configurations and different realizations of Gaussian 
 noise and then report a statistical summary. 
 By subjecting the obtained results to a two-component Gaussian mixture model algorithm, we derive the probability 
 distribution function (pdf) characterizing the spatial distribution of sky area in relation to sensitivity 
 range~\cite{scikit-learn}. We chose the two-component model as it proved to be the best to mirror the distribution of
 the underlying data, independently of the latter's features. Employing this estimated pdf, we generate 20,000 
 samples of sky area and range sensitivity values. Subsequently, we apply a filtering criterion, retaining samples 
 falling within the 5th and 95th percentiles and featuring positive (physical) values for both coordinates. Lastly, 
 we classify data points into bins based on the sensitivity range values, aligning the bin edges with integer values. 
 When varying \kagra's sensitivity range, we will employ unitary bin-widths, while when varying Virgo's 
 sensitivity range, we use quinary band-widths. 

\section{Results of Zero Spin Binary Black Hole Simulations}
\label{sec:bbh}
\begin{figure*}[!t]
\centering
\subfloat[\label{fig:HLK_bbh}]{%
     \includegraphics[width=0.46\textwidth]{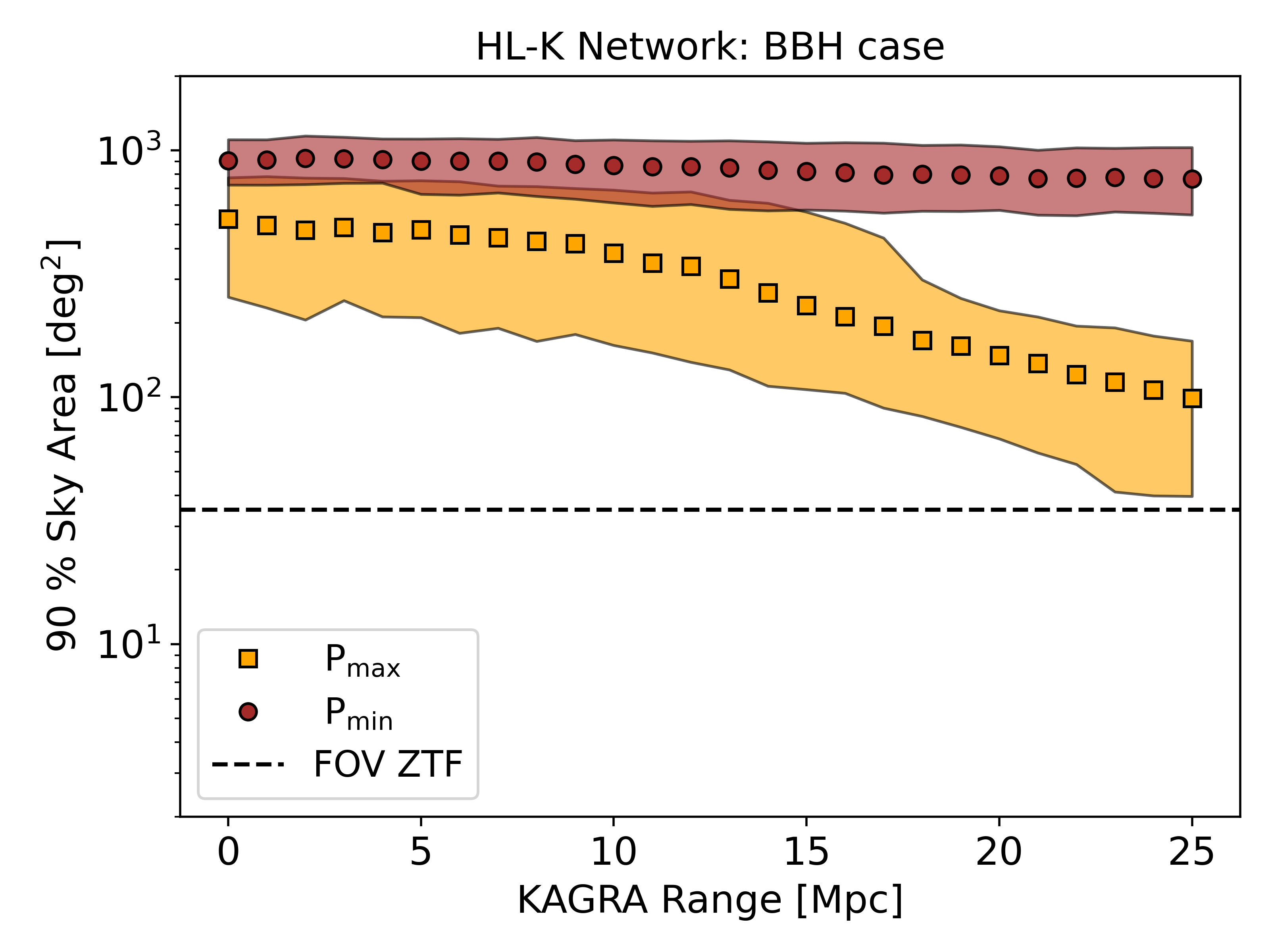}}%
\hfil
\subfloat[\label{fig:LK_bbh}]{%
   \includegraphics[width=0.46\textwidth]{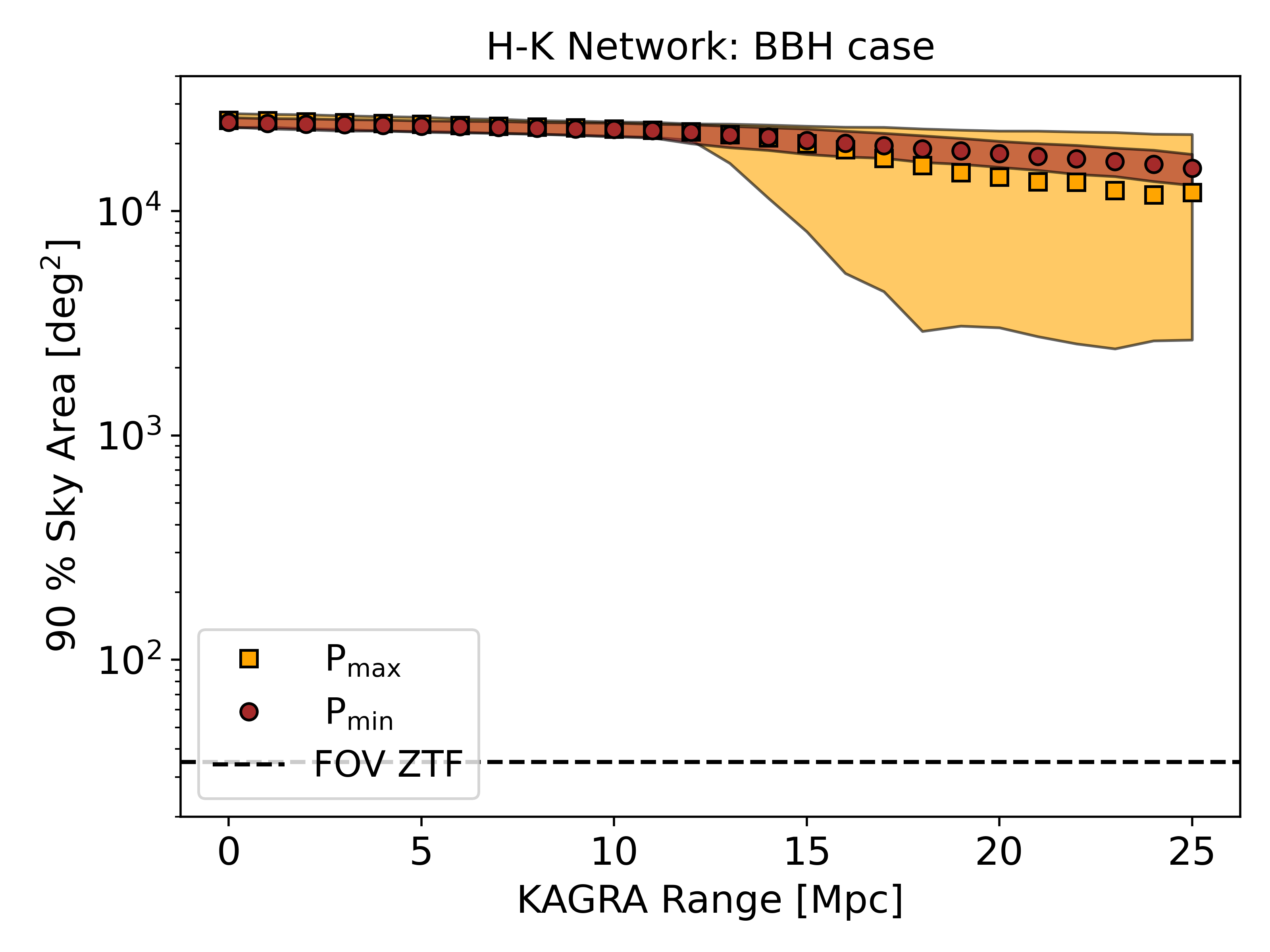}}%
\vskip\floatsep

\subfloat[\label{fig:HLV_bbh}]{
    \includegraphics[clip,width=0.46\textwidth]{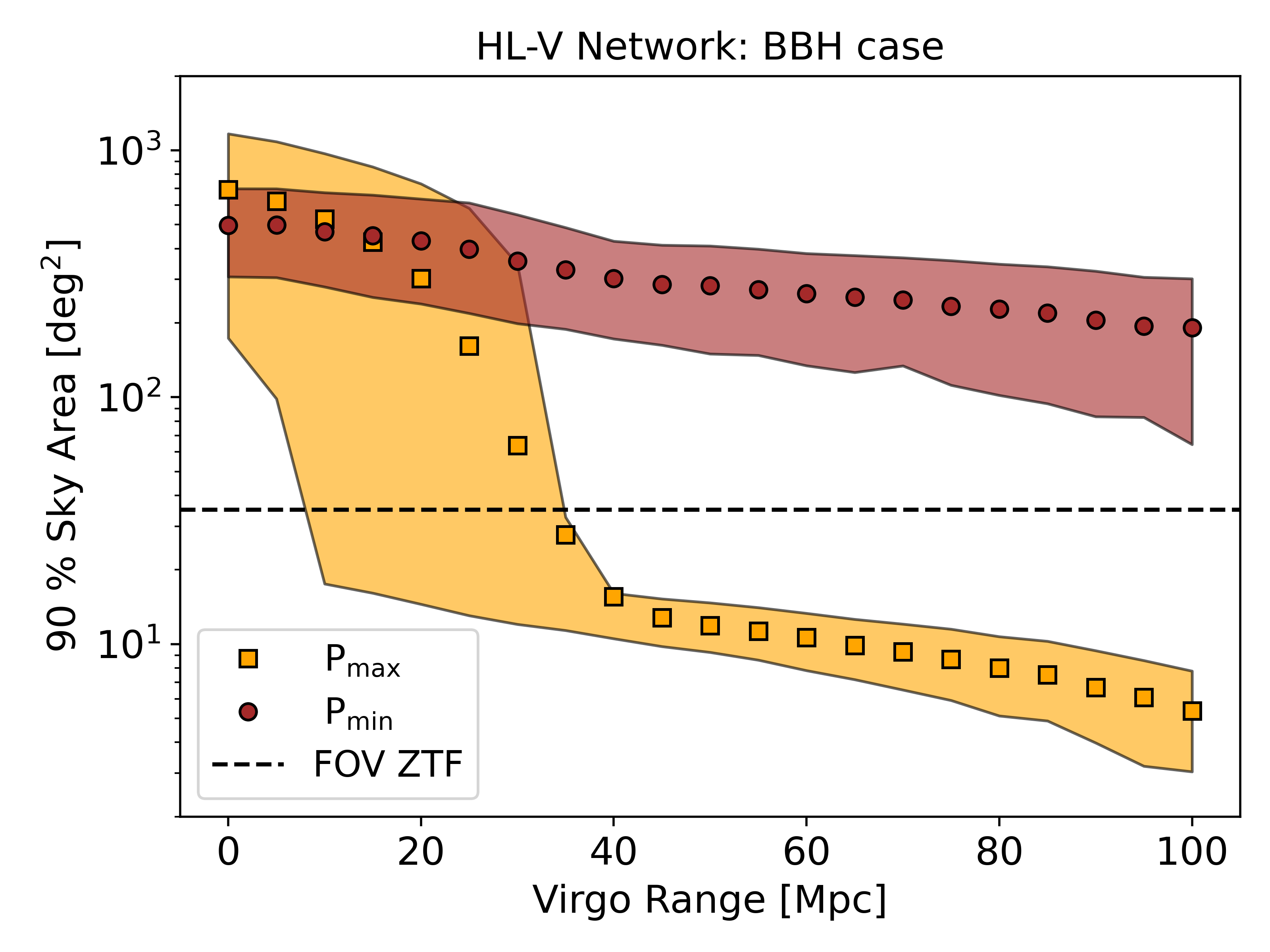}}%
\subfloat[\label{fig:LV_bbh}]{
    \includegraphics[clip,width=0.46\textwidth]{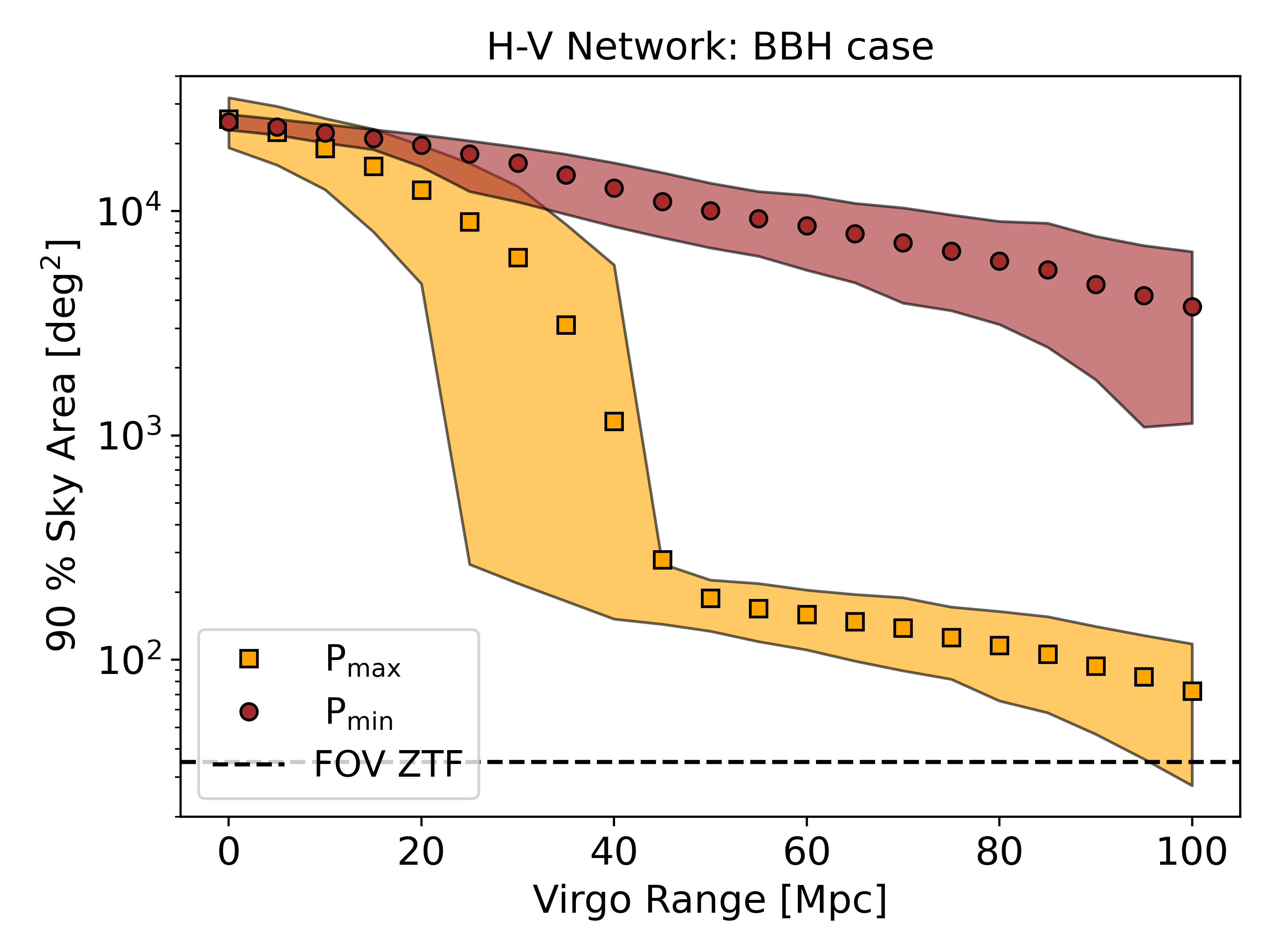}}%
\vskip\floatsep    
\subfloat[\label{fig:HLVK_bbh}]{
    \includegraphics[clip,width=0.46\textwidth]{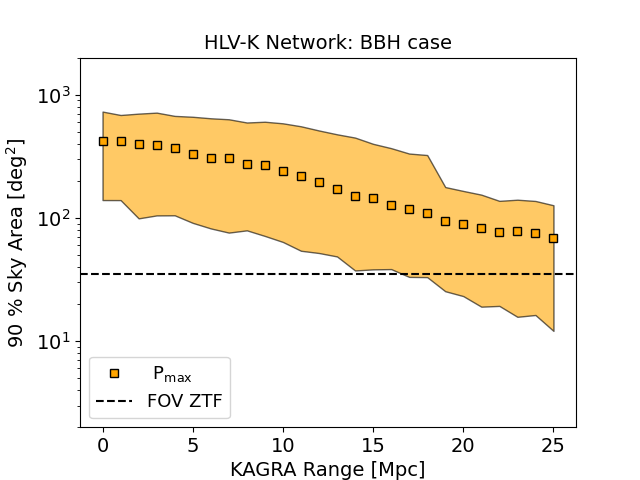}}%
\sbox0{%
\begin{minipage}[b]{0.46\textwidth}%
\mycaption{Zero-Spin Binary Black Hole Simulations}{ \RaggedRight \justifying The mean and the 90 \% probability contour of the sky area for varying detector range sensitivity values for a gravitational wave signal emitted 
by a binary black hole coalescence from the P$_{\Pmax}$ and P$_{\Pmin}$ source location for 
the least sensitive detector, brown and orange points and areas respectively. The points, brown dots, and orange squares, 
represent the mean value, 
and the areas stretch from the 5\% to the 95\% quantiles for each sensitivity range bin. The black dotted line corresponds to the field of view of the Zwicky Transient Facility. We obtained the 0 Mpc 
points from simulations excluding the least sensitive detector. We fixed the range sensitivity of the two LIGO 
detectors at 180 Mpc. \textbf{(a)}: Varying the \kagra~detector and employing the LHK network. \textbf{(b)}: 
Varying the \kagra~detector and employing the HK network. \textbf{(c)}: Varying the Virgo detector and employing 
the LHV network. \textbf{(d)}: Varying the Virgo detector and employing the HV network. \textbf{(e)}: Varying the 
\kagra~detector and employing the LHVK network. Virgo's sensitivity is fixed at 30 Mpc.  }
\label{fig:BBH}
\end{minipage}
}%
\hfil
\usebox0
\end{figure*}

 We simulate gravitational wave signals from non-spinning binary black hole mergers using the IMRPhenomXPHM 
 waveform approximant for both the signal's simulation and the likelihood 
 evaluation~\cite{Pratten:2020ceb}. We simulate equal-mass black holes with a detector-frame chirp 
 mass $\mathcal{M}$ of 50~M$_\odot$, where
\begin{equation}
\mathcal{M}=\frac{(m_1 m_2)^{3/5}}{(m_1 + m_2)^{1/5} } ,
\label{eq:chirp_mass}    
\end{equation}
with m$_{1/2}$ denoting the detector-frame mass of the two compact objects.
The source is placed at a distance of 2000 Mpc from us. The complete list of parameter values can be found 
in the first column of Table~\ref{tab:simulation} in \cref{sec: app_tables}. Following the procedure outlined in 
Sec.~\ref{sec:Method}, we perform six simulations and full parameter estimation runs for each sensitivity 
range value, and post-process the results into sky localization area values using a Gaussian Mixture model. 
The results are presented in Figure~\ref{fig:BBH}.

In Figure~\ref{fig:HLK_bbh}, we show the evolution of the mean and 90~\% probability contours of the sky area, 
dots and squares, and shaded area, respectively, for the HL-K detector network varying the range sensitivity of the 
\kagra~detector only. Henceforth, we will separate with a dash the letter referring to the detector of the 
network whose range sensitivity we have varied for the specific plot. The brown and orange points and 
areas were obtained by placing the source in the P$_{\Pmax}$ and P$_{\Pmin}$ source location, respectively, 
for the \kagra~detector (see Table~\ref{tab:localization}). The black dotted line corresponds to the field of view of the Zwicky Transient Facility. By 0 Mpc we refer to the results obtained with the 
two LIGO detectors only at a fixed sensitivity range of 180 Mpc. For all the following sky area plots, we computed 
the 0 Mpc value with the same detector network used for the other sensitivity values but excluding the detector 
whose sensitivity range is varied. For Figure~\ref{fig:HLK_bbh}, while in the P$_{\Pmax}$ location case, the mean 
sky area value decreases notably, when increasing \kagra's sensitivity range, it remains roughly constant in the 
P$_{\Pmin}$ location case. The decrease of the mean sky area is most significant from $\sim$10 Mpc on-wards.  

Figure~\ref{fig:LK_bbh} shows the sky area results for the P$_{\Pmax}$ and P$_{\Pmin}$ source locations 
for the H--K detector network varying the \kagra~sensitivity range. For the P$_{\Pmax}$ case, the sky area 
improves considerably after $\sim$15~Mpc, with the 90\% quantiles spanning to one order of magnitude lower 
values than in the P$_{\Pmin}$ case. Also the latter improves slightly from 15~Mpc on-wards.
The mean sky area values for the P$_{\Pmax}$ case are of the order $1.5\times10^4$~deg$^2$, 
when \kagra~has a 25~Mpc sensitivity range, compared to $2\times10^4$~deg$^2$ for the P$_{\Pmin}$ case. 
The overall improvement is not significant for precisely localizing the source, and the sky area values are 
two orders of magnitude higher than for the corresponding three-detector network HL--K.

In Figure~\ref{fig:HLV_bbh}, we report the same quantities as in Figure~\ref{fig:HLK_bbh}, but for 
the HL--V detector network, using the P$_{\Pmax}$ and P$_{\Pmin}$ source location for the Virgo detector. 
We can see a clear difference between the P$_{\Pmax}$ and P$_{\Pmin}$ distributions, with the decrease in 
sky area being almost two orders of magnitude more in the former source location case. We notice how the Virgo 
detector with a sensitivity range of 35 Mpc and above would provide a significant improvement, reducing the mean 
sky area to $\sim$10~Mpc. With a Virgo detector range sensitivity between 25 and 35~Mpc, the inferred sky area 
improves considerably, although it exhibits a high variability. In the P$_{\Pmin}$ case, the improvement in 
sky area is less remarkable as the mean changes from $\sim$600~deg$^2$ with the HL detector network to 
$\sim$250~deg$^2$ for the LH--V network when Virgo has a sensitivity range of 100~Mpc.

The corresponding two-detector network, H--V, results are shown in Figure~\ref{fig:LV_bbh}. 
We notice that both the P$_{\Pmax}$ and P$_{\Pmin}$ distributions follow a similar trend to the ones 
obtained for the HL--V detector network. In the P$_{\Pmax}$ case, we see the most significant improvement 
in sky area for sensitivity range values between 25 and 45~Mpc. In this range, the mean sky area value drops 
from $\sim$10$^4$ to $\sim$200~deg$^2$. From 45~Mpc onwards, the betterment of the mean sky area value is not 
as meaningful, reaching $\sim$80~deg$^2$ when Virgo has a sensitivity range of 100 Mpc. In the P$_{\Pmin}$ case, 
the distribution of the mean sky area values is quasi-linear, progressing continuously from $\sim$2$\times$10$^4$ 
to 400~deg$^2$ for 0 and 100~Mpc respectively.

Finally, Figure~\ref{fig:HLVK_bbh} shows the mean sky area values and 90\% probability contours obtained 
with the methods outlined in Sec.~\ref{sec:post_proc} from simulated detections of binary black hole merger 
signals with the HLV-K detector network when varying the sensitivity range of the \kagra~detector only. 
We fix the sensitivity of the Virgo detector at 30~Mpc and employ the P$_{\Pmax}$ source location for the 
\kagra~detector in all the runs. We observe a continuous improvement of the sky area from 5 to 25~Mpc, 
with the mean value passing from $\sim$500~deg$^2$ to $\sim$100~deg$^2$. This suggests that even in a four-detector 
network, adding an interferometer up to ten times less sensitive than the others but at a different spatial location
and with a different relative orientation leads to a 75\% reduction in the mean 
sky localization area inferred with the network.

\section{Results of Binary Neutron Star Simulations}
\label{sec:bns}

\begin{figure*}[t!]
\centering
\subfloat[\label{fig:HLK_ns}]{%
     \includegraphics[width=0.46\textwidth]{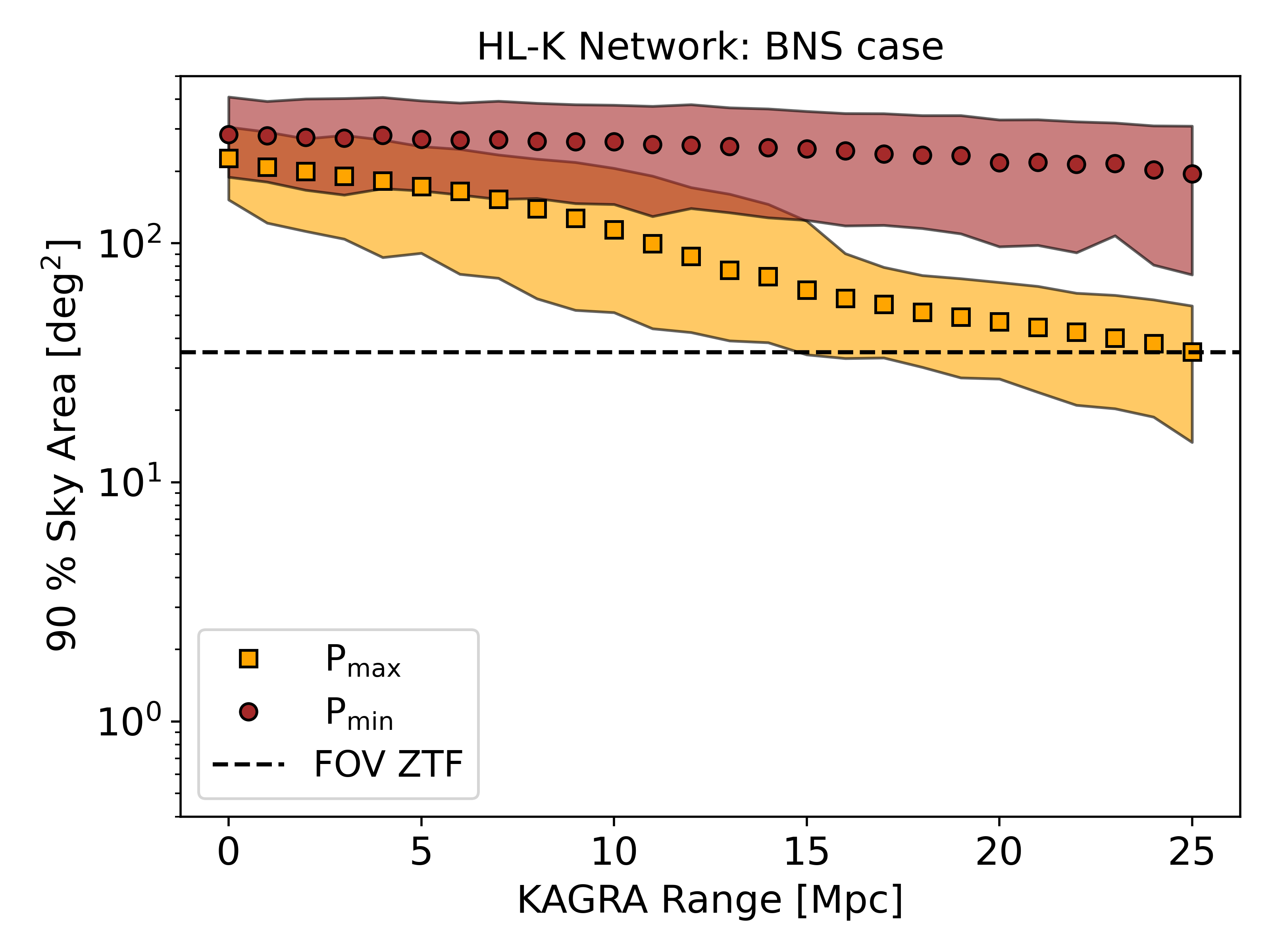}}%
\hfil
\subfloat[\label{fig:LK_ns}]{%
   \includegraphics[width=0.46\textwidth]{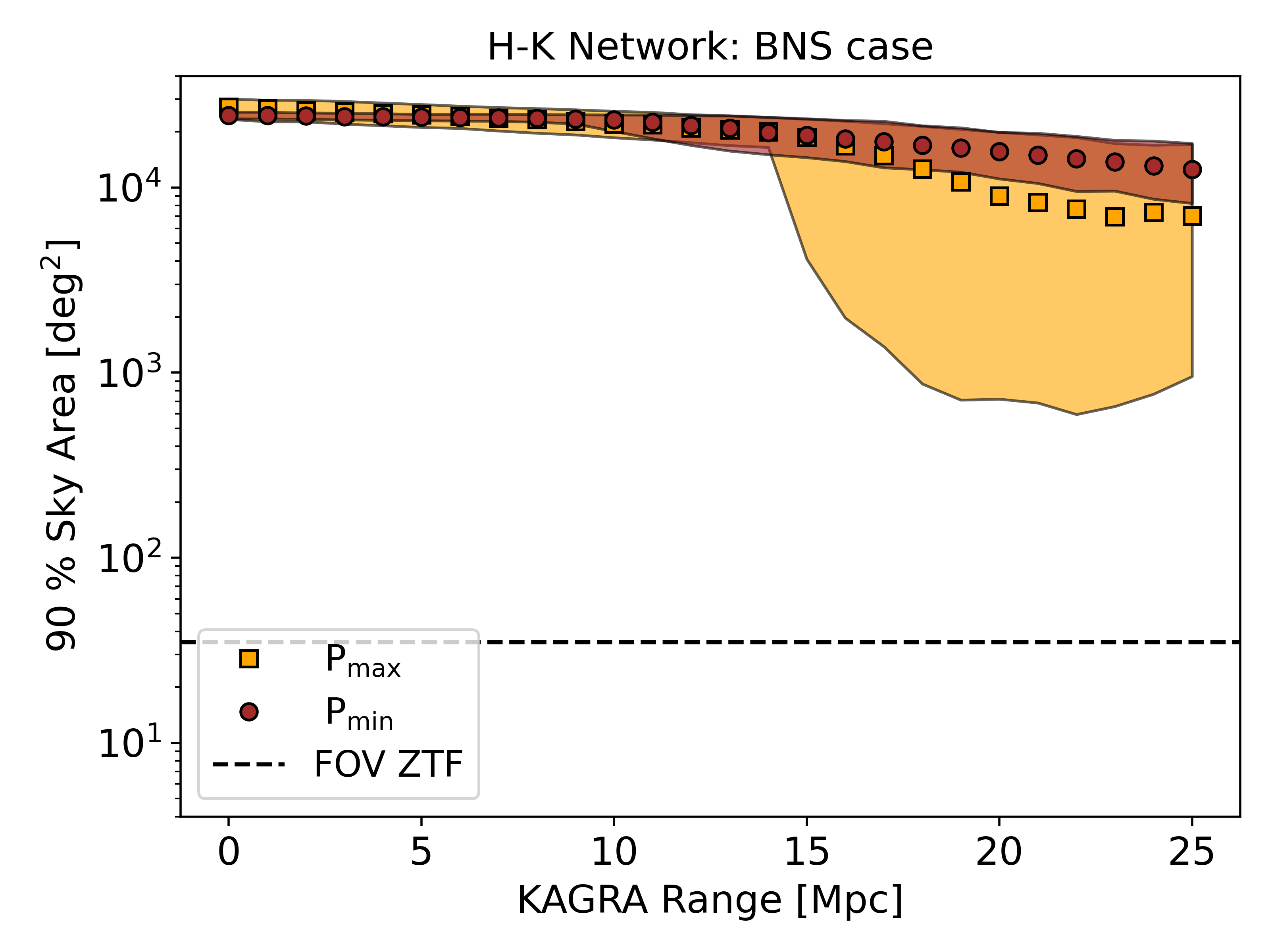}}%
\vskip\floatsep

\subfloat[\label{fig:HLV_ns}]{
    \includegraphics[clip,width=0.46\textwidth]{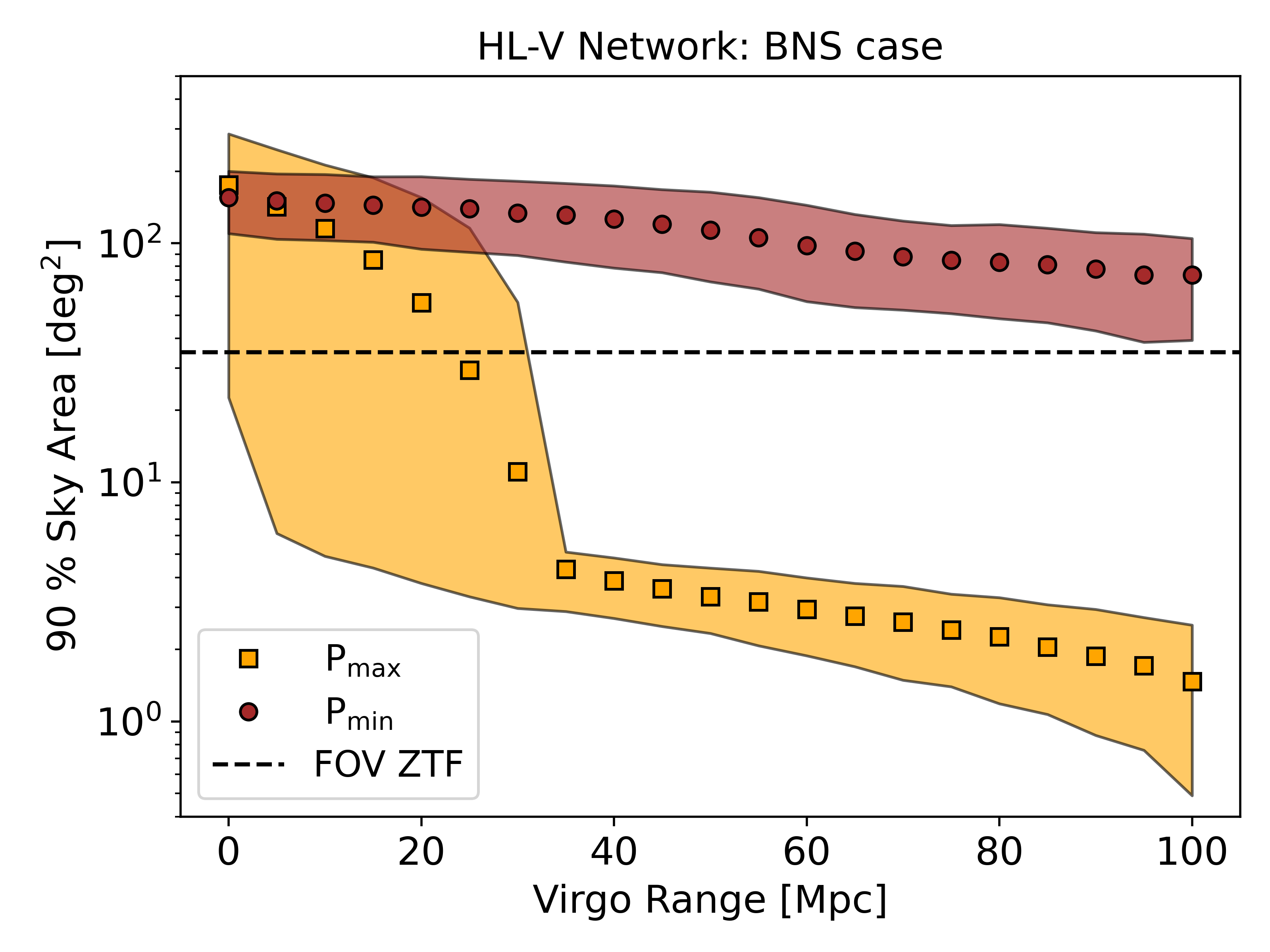}}%
\subfloat[\label{fig:LV_ns}]{
    \includegraphics[clip,width=0.46\textwidth]{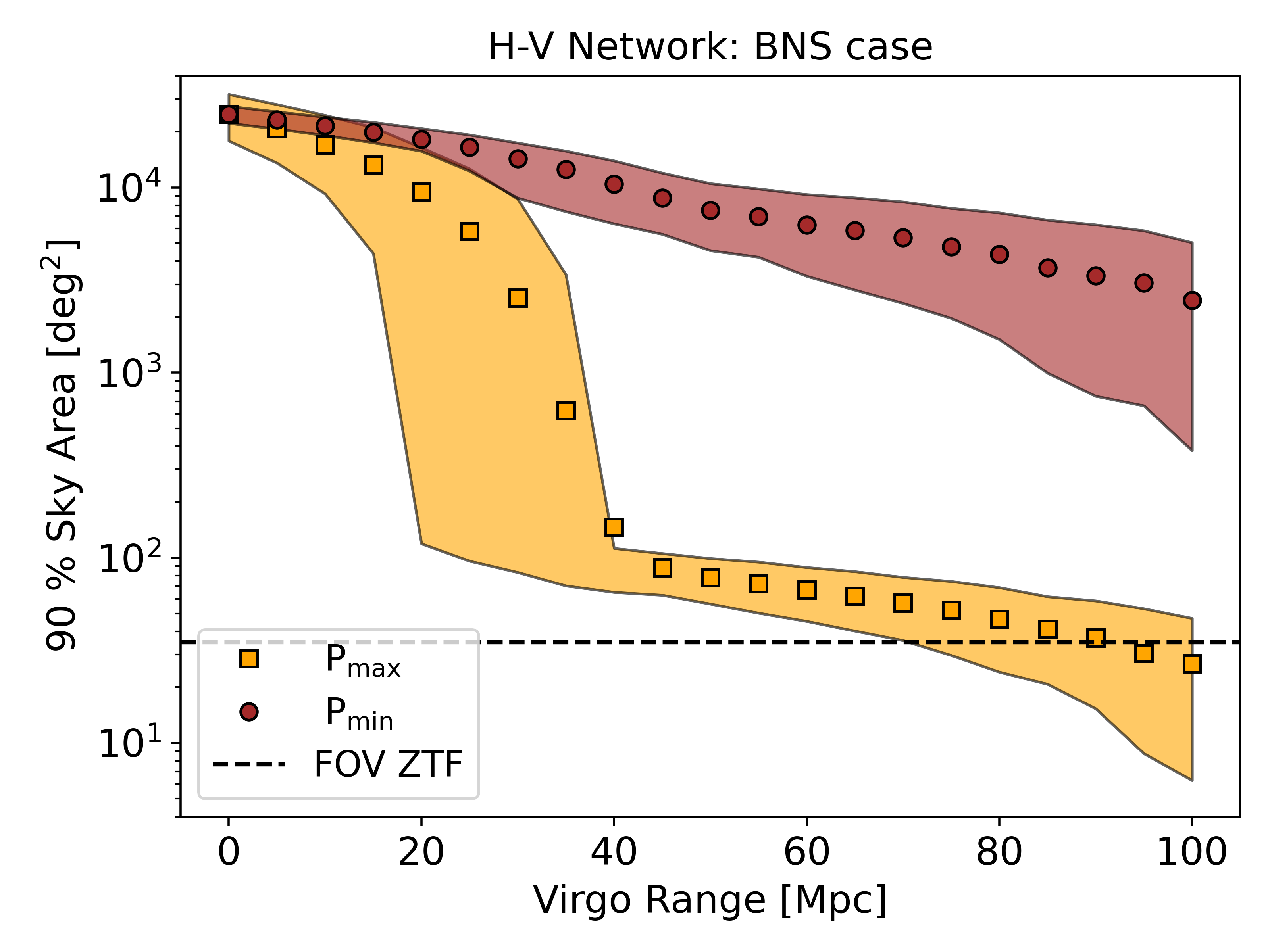}}%
\vskip\floatsep    
\subfloat[\label{fig:HLVK_ns}]{
    \includegraphics[clip,width=0.46\textwidth]{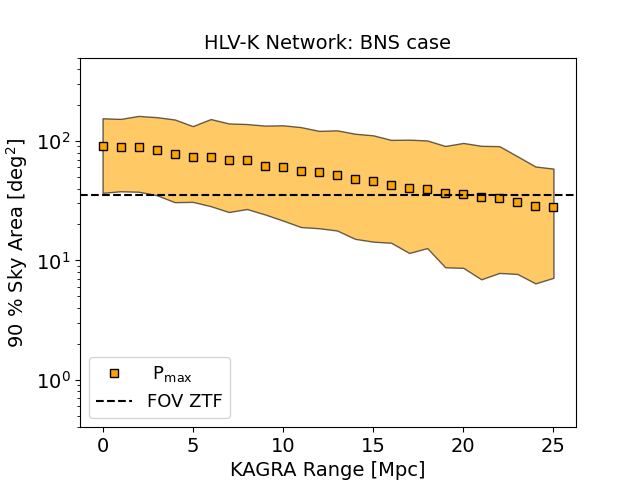}}%
\sbox0{%
\begin{minipage}[b]{0.46\textwidth}%
\mycaption{Binary Neutron Star Simulations}{ \RaggedRight \justifying The mean and the 90\% probability contour 
of the sky area for varying detector range sensitivity values for a gravitational wave signal emitted by a 
binary neutron star coalescence from the P$_{\Pmax}$ and P$_{\Pmin}$ source location for 
the least sensitive detector, brown and orange dots, and areas, respectively. The dots represent the mean value, 
and the areas stretch from the 5\% to the 95\% quantiles for each sensitivity range bin. The black dotted line corresponds to the field of view of the Zwicky Transient Facility. We obtained the 0~Mpc 
points from simulations excluding the least sensitive detector. We fixed the range sensitivity of the two LIGO 
detectors at 180~Mpc. \textbf{(a)}: Varying the \kagra~detector and employing the LHK network. \textbf{(b)}: 
Varying the \kagra~detector and employing the HK network. \textbf{(c)}: Varying the Virgo detector and employing 
the LHV network. \textbf{(d)}: Varying the Virgo detector and employing the HV network. \textbf{(e)}: Varying 
the \kagra~detector and employing the LHVK network. Virgo's sensitivity is fixed at 30~Mpc.}
\label{fig:BNS}
\end{minipage}
}%
\hfil
\usebox0
\end{figure*}

We simulate gravitational wave signals from non-spinning binary black hole merger following the same procedure of Sec.~\ref{sec:bbh}. 
The binary neutron stars are placed at a distance
from Earth of 100~Mpc and have a chirp mass of $1.198$~M$_\odot$, comparable to the first binary neutron star 
merger event GW170817~\cite{TheLIGOScientific:2017qsa}. The complete list of parameter values can be found in the
second column of Table~\ref{tab:simulation} in \cref{sec: app_tables}. For the gravitational wave signal simulation and 
parameter estimation, we use the IMRPhenomPv2~\cite{Hannam:2013oca} waveform approximant, and to speed up 
the evaluation of the likelihood function, we employ the reduced order quadrature basis from the analysis of 
GW170817~\cite{Canizares:2013ywa}. We list the priors in the second column of Table~\ref{tab:prior}. 
The results are presented in Figure~\ref{fig:BNS}.

In Figure~\ref{fig:HLK_ns}, we show the mean and the 90\% quantile region of the sky localization area as 
a function of the sensitivity range of the \kagra~detector for the P$_{\Pmax}$ and P$_{\Pmin}$ source locations 
for \kagra. We use the same color scheme as in Sec.~\ref{sec:bbh}. The points and regions have been obtained 
with the methods detailed in Sec.~\ref{sec:post_proc} from the parameter estimation results of simulated 
gravitational wave signals emitted by the coalescence of binary neutron star mergers using the HL--K detector 
network. The 0~Mpc point corresponds to the sky area obtained from an HL detector network run with identical 
source locations and parameters. While in the P$_{\Pmin}$ source location case, there is no improvement in the 
mean sky area value from 0 to 25~Mpc, in the P$_{\Pmax}$ case, the value decreases continuously from 
$\sim$200~deg$^2$ to $\sim$30~deg$^2$ for the same range values.

Figure~\ref{fig:LK_ns} shows the sky area results for the P$_{\Pmax}$ and P$_{\Pmin}$ source locations 
for the H--K detector network, varying sensitivity range of \kagra. For the P$_{\Pmax}$ case, the sky area 
improves from $\sim$15~Mpc onwards, with the 90\% quantiles spanning to one order of magnitude lower values 
than in the P$_{\Pmin}$ case. The mean sky area values for the P$_{\Pmax}$ case are of the order 
$1.2\times10^4$~deg$^2$, with \kagra~having 25~Mpc range sensitivity, compared to $1.5\times10^4$~deg$^2$ for 
the P$_{\Pmin}$ case. This trend is coherent with the corresponding binary black hole case shown in 
Figure~\ref{fig:LK_bbh}.

Figure~\ref{fig:HLV_ns} shows the mean sky area values and the 90\% probability areas obtained from the 
parameter estimation runs using the HL--V detector network placing the source in the P$_{\Pmax}$ and P$_{\Pmin}$ 
source locations for Virgo. As in Figure~\ref{fig:HLV_bbh}, we see that for the P$_{\Pmin}$ source location, the 
sky area values do not improve when Virgo's sensitivity range is wider, while in the P$_{\Pmax}$ case, there is 
an improvement starting from 15~Mpc. This improvement is significant up to $\sim$40~Mpc, going from 
$\sim$200~deg$^2$ without Virgo to $\sim$4~deg$^2$ when Virgo has a range sensitivity of 40~Mpc. 
The sky area values remain roughly constant from 40~Mpc to 100~Mpc for this optimal case. In the P$_{\Pmin}$ case, 
the sky area does not change between 0 and 40~Mpc and decreases minimally from 40~Mpc on-wards.

As for the \ac{BBH} case, the sky area for the two detector networks, H--V, exhibits a similar evolution 
for the sensitivity range of Virgo. These results are visible in Figure~\ref{fig:LV_ns}. 
In the optimal case, we see a considerable refinement of the sky area between 25 and 40~Mpc, with 
the mean value passing from 10$^4$ to 10$^2$~deg$^2$. This value continues to improve continuously
up to 30~deg$^2$ at 100~Mpc. Also in the P$_{\Pmin}$ case, differently from the correspondent three 
detector network results in Figure~\ref{fig:HLV_ns}, we see a noticeable improvement with increasing 
sensitivity range. Indeed, the mean sky area value is reduced to 3$\times$10$^3$~deg$^2$ at 100~Mpc, 
with the 90\% area reaching down to 400~deg$^2$, from 2.5$\times$10$^4$ at 0~Mpc.

Lastly, Figure~\ref{fig:HLVK_ns} is the correspondent of Figure~\ref{fig:HLVK_bbh} for simulated 
detections of binary neutron star merger signals with the HLV--K detector network when varying the 
sensitivity range of the \kagra~detector only. The network setup in the simulations is identical to the binary 
black hole case. Similarly, we observe a continuous improvement of the sky area, with the mean value 
decreasing from $\sim$100~deg$^2$ to $\sim$30~deg$^2$. The 90\% quantile region follows a similar trend, 
with the lower end going from $\sim$40~deg$^2$ without \kagra~to $\sim$5~deg$^2$ when \kagra~reaches a 
sensitivity range of 25~Mpc.

\section{Results of highly-spinning binary black holes}
\label{sec:spinning}
Finally, we simulate gravitational wave signals from 
highly spinning aligned binary black holes employing the Hanford-Livingston-Virgo detector network and 
perform parameter estimation varying the range sensitivity of the Virgo detector only. 
In contrast to the previous sub-section, we aim to understand how the sensitivity of the Virgo detector impacts the inference of the intrinsic properties of the source. We focus on a high-spin binary
black hole merger as the spin-related parameters, e.g., the effective spin and the individual spin magnitudes, 
are only broadly constrained 
by current detector network configurations, while they encode a large amount of information. Tighter constraints on 
these variables could lead to an enhanced understanding of the 
Universe's black hole population properties and their formation channels~\citep{Bavera:2020inc}.  

We simulate a signal with the following arbitrary choice of parameters: we employ a chirp mass of 27.93 M$_\odot$ and a $\chi_{\rm eff}$ value of 0.91, and assume that the spins are aligned
so that $\chi_p=0$ and the tilt angles are null. The complete list of values is reported in the third column of
Table~\ref{tab:simulation} in \cref{sec: app_tables}, while the priors used for the runs are listed in the third column
of Table~\ref{tab:prior}. We also adjust the settings for the \dynesty sampler, as visible in Table~\ref{tab:dinesty}
in \cref{sec: app_tables}, to reduce the bias in the posterior probability results.  

The analysis of these simulations takes significant computational effort. 
To alleviate this, we will apply a rejection sampling approach using a base analysis as a proposal distribution and reweighting for each rescaled Virgo sensitivity.
Specifically, we begin with the posterior obtained for the HLV network when Virgo has a sensitivity of 50 Mpc and apply the standard Nested Sampling algorithm as described in \cref{sec:Method}.
We choose 50 Mpc, as this is above the point where the extrinsic parameters are well constrained by triangulation.
This initial HLV analysis takes $\approx 8$~hrs using 30 CPU cores.
Each reweighting takes $\approx 10$~minutes on a single CPU core.
This vast improvement is only possible because as we increase the sensitivity of Virgo, the posterior narrows marginally such that the base analysis is an excellent generating distribution.

After performing the base analysis and obtaining a set of posterior samples $\{\theta_i\}$, we obtain the weight for the $i$th sample as
\begin{equation}
    W_i=\frac{f(\theta_i)}{Mg(\theta_i)}, \quad \text{with} \quad  f(\theta_i) \leq M \times g(\theta_i),
    \label{eq:weight}
\end{equation}
where $g(\theta_i)$ is the generating distribution, $f(\theta_i)$ is the resampling distribution, and $M$ is a factor used to ensure that
the generating distribution encompasses the resampling distribution, effectively normalizing the weights to the range $[0,1)$.  
In our case, the generating distribution is $g(\theta_i)=\mathcal{L}(\theta_i|\rm{H_{180}})\mathcal{L}(\theta_i|\rm{L_{180}})\mathcal{L}(\theta_i|\rm{V_{50}})\pi(\theta_i)$,
while the resampling distribution
is $f(\theta_i)=\mathcal{L}(\theta_i|\rm{H_{180}})\mathcal{L}(\theta_i|\rm{L_{180}})\mathcal{L}(\theta_i|\rm{V_{j}})\pi(\theta_i)$ where the subscripts for the detectors indicate the sensitivity range, expressed in Mpc, of the PSDs used for the evaluation of the noise realizations
and $j$ ranges from 60 to 180~Mpc. Therefore, since the likelihood for the Hanford and Livingston data is identical, we can simplify the weights from Eq.~\ref{eq:weight} to be:
\begin{equation}
    W_i=\frac{\mathcal{L}(\theta_i|\rm{V_{j}})}{M\mathcal{L}(\theta_i|\rm{V_{50}})}.
    \label{eq:weight_adapted}
\end{equation}
The basic idea of rejection sampling is to draw random samples from the generating distribution $g$, 
calculate the related weights, and compare these to a random number taken from a uniform distribution between 
0 and 1~\citep{Casella:2004}. If the weight is bigger than the random number, we accept the sample and include it
in the posterior, otherwise, we discard it. It is important to carefully choose the value of $M$ so that it 
satisfies the condition in Eq.~\ref{eq:weight} and at the same time does not make the sampling inefficient by 
being too large. We choose the value of $M$ by calculating
\begin{equation}
    \log(M)=\mathrm{Max}_i\left(\frac{\log(\mathcal{L}(\theta_i|\mathrm{V_{j}}))}{\log(\mathcal{L}(\theta_i|\mathrm{V_{50}}))}\right)+0.1\,,
    \label{eq:factor_M}   
\end{equation}
over set of samples.

To account for the high variability of the individual noise realizations when using the same PSD and the resulting 
oscillating efficiencies,
for each sensitivity range value we perform 20 iterations of the resampling routine. We focus on the sensitivity range 
between 50 and 180~Mpc and 
employ PSDs obtained with the methods described in Section~\ref{sec:Method} progressively increasing the sensitivity by tens. We then 
combine the samples from the multiple iterations before post-processing the data. 

\begin{figure*}[t!]
\centering
\subfloat[\label{fig:chirp_mass}]{%
     \includegraphics[width=0.49\textwidth]{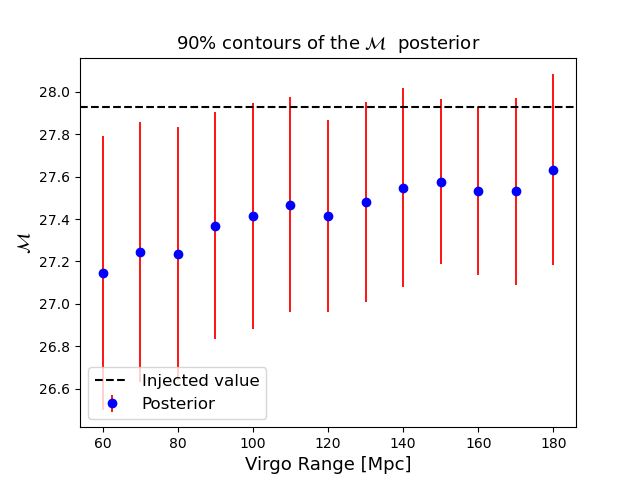}}%
\hfil
\subfloat[\label{fig:mass_ratio}]{%
   \includegraphics[width=0.49\textwidth]{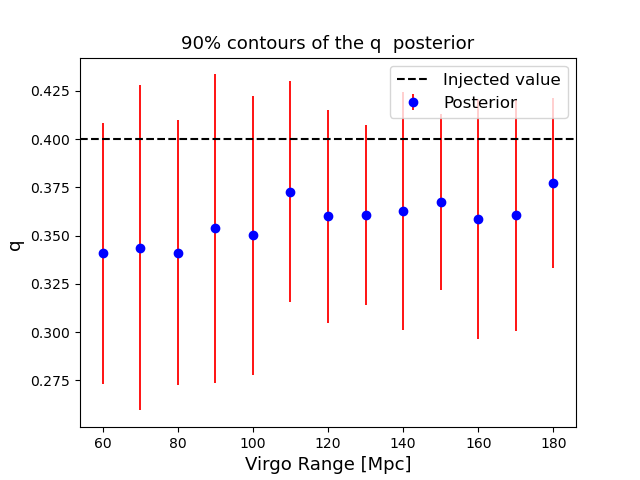}}%
\vskip\floatsep

\subfloat[\label{fig:chi_eff}]{
    \includegraphics[clip,width=0.49\textwidth]{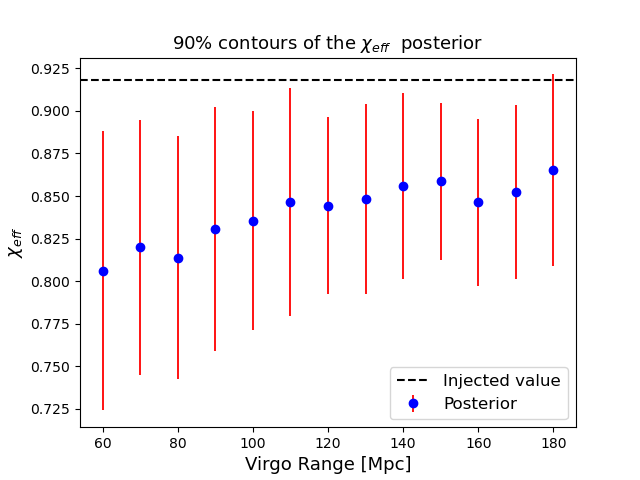}}%
\subfloat[\label{fig:chi_p}]{
    \includegraphics[clip,width=0.49\textwidth]{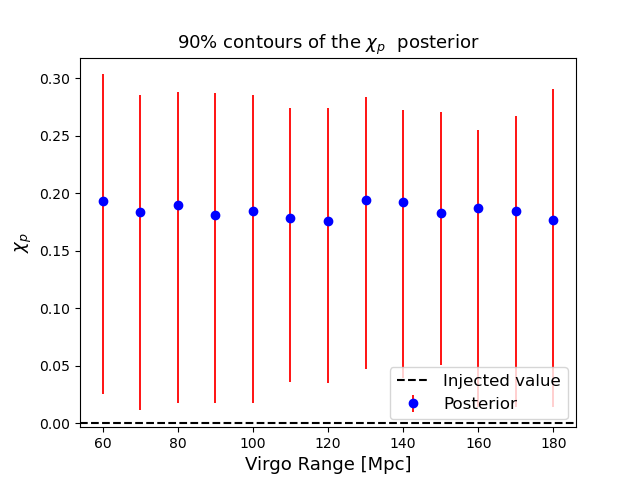}}%
\vskip\floatsep    
\subfloat[\label{fig:ra}]{
    \includegraphics[clip,width=0.49\textwidth]{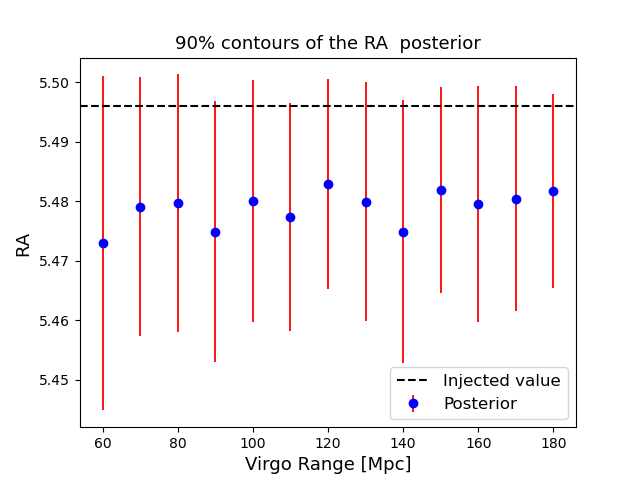}}%
\sbox0{%
\begin{minipage}[b]{0.49\textwidth}%
\mycaption{High-Spin Binary Black Hole Simulations}{ \RaggedRight \justifying The mean, blue dots, and 90\% posterior probability 
contour, red line, of the detector-frame chirp mass obtained varying the sensitivity range of Virgo in the parameter estimation analysis of a 
simulated high-spin binary black hole merger signal. We employ the HL-V detector network, fixing the sensitivity range, 
noise realization and detector response of the two LIGO detectors. The black dotted line represents the true chirp mass value of 
the simulated signal.}
\label{fig:BBH_spin}
\vspace{30mm}
\end{minipage}
}%
\hfil
\usebox0
\end{figure*}

In Figure~\ref{fig:BBH_spin} we show a summary of our results. In each of the panels, we show the mean, blue dots, and 90\% 
probability contours, red lines,
of the selected parameter versus the different sensitivity ranges of the Virgo detector. 
The 90\% contour is centered around the mean for 
all variables except for $\chi_{\rm{p}}$. To account for the latter's true value corresponding to the boundary of the 
parameter's lower constraint, 
its 90\% contour is taken from the lower boundary to the 90\% quantile.
The true values used for the simulation of the gravitational wave signal correspond to the black dotted horizontal lines. 
Overall, we notice a consistent improvement in the mean value and a tightening of the 90\% probability contours when 
increasing Virgo's sensitivity.
This trend is particularly evident in panels~\ref{fig:chirp_mass} and~\ref{fig:chi_eff}, depicting the results for the detector-frame 
chirp mass and
the effective spin parameter. The latter is a dimensionless parameter quantifying the alignment of the spins of the 
binary components are given by:
\begin{equation}
    \chi_{\rm{eff}}=\frac{(m_1 \cdot \chi_1+m_2 \cdot \chi_2)\cdot \hat{L} }{M},
\end{equation}
with $m_{1/2}$ denoting the masses of the two black holes, M their total mass, L their angular momentum and $\chi_{1/2}$ the 
individual dimensionless spins. For both, the chirp mass and effective spin, the mean value gets continuously closer to the true value
reaching a 60\% and 45\% improvement at 180~Mpc, respectively. At the same time, the 90\% probability contours decrease by 30\% for both
parameters from 60 to 180~Mpc. The mass ratio shown in panel~\ref{fig:mass_ratio} shows a similar trend going from 60 to 180~Mpc, 
with a 60\% improvement in the mean value and a 30\% decrease in the 90\% probability contour. We notice that in this case, 
the most significant betterments happen at sensitivities above 100~Mpc. Finally, in panels~\ref{fig:chi_p} and~\ref{fig:ra} we 
report the results for $\chi_p$ and the right ascension, respectively. The parameter $\chi_{p}$, introduced in~\citet{Schmidt:2012rh}, 
combines the spin magnitudes and their relative orientation to the total angular momentum, effectively characterizing the 
degree of precession in a binary black hole system. As we are analyzing a non-precessing binary system, 
we expectedly recover $\chi_p$ to be consistent with zero at all sensitivities. For the right ascension, we obtain 
only minor improvements passing from 60 to 180~Mpc. This is not surprising because, as noted in Section~\ref{sec:bbh}, the 
sky-localization shows the largest improvements at sensitivities between 20 and 40~Mpc. Overall, we can conclude that when Virgo reaches
sensitivities starting from approximately half of the LIGO sensitivities, i.e. 90/100~Mpc, the inference of the intrinsic source 
parameters improves significantly.

\section{Minimum frequency study}
\label{sec:min_freq}
Gravitational wave signals from merging black holes and neutron stars sweep up in frequency during the inspiral stage before their eventual merger 
and ringdown.
From a simple post-Newtonian calculation, it can be shown (see, e.g. \citet{Peters_1963}) that the time-to-merger from a frequency $f_0$ is
\begin{equation}
    t_m = 1.5\, \textrm{s}
    \left( \frac{20 \, M_{\odot}}{\mathcal{M}} \right)^{5/3}
    \left( \frac{20 \, \textrm{Hz}}{f_0} \right)^{8/3}\,,
    \label{eqn:tm}
\end{equation}
where $\mathcal{M}$ is the detector-frame chirp mass defined in Eq.~\ref{eq:chirp_mass}.
The power of $8/3$ means that signals spend far longer at lower frequencies than at the higher frequencies near the merger 
(this expression breaks down near the merger itself but is a good approximation for the early inspiral).
However, the noise curve of the detector steeply rises at low frequencies due to seismic noise.
As such, analyses often assume a minimum frequency $f_{\rm min}$ below which the signal has negligible contributions.

Specifically, in standard approaches to Bayesian inference for signals, we assume a stationary Gaussian noise resulting in a log-likelihood:
\begin{equation}
    \log \mathcal{L}(d|\theta) \propto \sum_{k}\frac{2|\tilde{d}_{k} - \tilde{\mu}_{k}(\theta)|^2}{T S_k}\,,
\end{equation}
where $T$ is the data duration, $k$ is the index of the frequency bin, $\tilde{d}$ is the complex frequency-domain data, 
$\tilde{\mu}(\theta)$ is the frequency-domain source model evaluated at a set of source parameters $\theta$, and $S$ is 
the Power Spectral Density (see, e.g., \citet{Finn:1992wt, Veitch:2014wba}).
In practice, the minimum frequency is implemented by neglecting to sum components below $k_{\rm min}$, the frequency bin 
associated with $f_{\rm min}$.

The introduction of $f_{\rm min}$ then informs the choice of $T$, the duration of data to analyze. Choosing $T$ to be much larger than 
the actual signal duration is inefficient, requiring unnecessary computation and ultimately increasing the computational cost 
of an analysis.
The standard in the field remains to take $f_{\rm min}=20$~Hz and then choose $T$ to be the next power of two greater 
than the estimated time-to-merger (either using Eqn.~\ref{eqn:tm} or an improved model incorporating other physical effects).

This standard has been in place since the first detection \citep{TheLIGOScientific:2016wfe}, though with some exceptions. 
It is common to increase $f_{\rm min}$ as a mechanism to mitigate non-Gaussianity in the detector \citep{Davis:2022ird}.
However, it has also been decreased to maximize the number of cycles in-band for analysis 
(see, e.g., \citet{LIGOScientific:2020aai} and \citet{LIGOScientific:2020ufj} which used a minimum frequency of 19.4~Hz and 11~Hz, respectively).  

While the $20$~Hz rule of thumb is built on a solid understanding of the detector performance over the first observing runs, 
recent upgrades have seen improvements in the low-frequency sensitivity and future detector developments aim to improve performance further.
This leads us to ask: what conditions should be met for $f_{\rm min}$ to be decreased? Our ultimate goal is to develop a simple-to-compute
metric that can provide gravitational-wave astronomers with a better-informed rule of thumb for choosing $f_{\rm min}$.

To construct our metric, we use the matched-filter Signal-to-Noise-Ratio (SNR)
\begin{equation}
    \rho_{\rm mf} = 
    \frac{\langle \tilde{d}, \tilde{\mu}(\theta)\rangle}{\sqrt{\langle \tilde{\mu}, \tilde{\mu}(\theta)\rangle}}\,,
    \label{eqn:mf}
\end{equation}
where $\langle a, b \rangle$ denotes the noise-weighted inner product:
\begin{equation}
   \langle a, b \rangle = \frac{4}{T}\sum_{k=k_{\rm min}}^{k_{\rm max}} \Re\left(\frac{a_k b_k^*}{S_k}\right)\,,
\end{equation}
and $k_{\rm min}$ corresponds to the minimum frequency bin $f_{\rm min}$ while $k_{\rm max}$ corresponds the maximum frequency bin 
(which we set to $2048$~Hz).

Taking time-domain data from a given detector, we Fourier transform to the frequency domain and add to it a simulated fiducial signal; 
we then use that same signal to calculate the matched-filter SNR. I.e., the simulated signal added to the data is also used in 
Eqn.~\ref{eqn:mf}.
 
Finally, we vary $f_{\rm min}$ and define $\hat{f}_{\rm min}$ to be the minimum frequency at which $\rho_{\rm mf}$ decreases 
by $\epsilon=0.1\%$ relative to the value as calculated at $f_{\rm min}=15$~Hz (an arbitrary choice considered to be sufficiently low 
to represent the maximum SNR).
Within this definition, there are two tuning parameters. First, the choice of fiducial signal. We apply two choices: a fiducial \ac{BBH} and a 
fiducial BNS.
For both signals, we use the \texttt{IMRPhenomXAS} waveform model \citep{Pratten:2020fqn} with zero-spin, equal masses, and 
arbitrary choices of $\theta_{\rm JN}=0.4$, $\psi=2.659$, $\phi=1.3$, $\alpha=1.375$, $\delta=-1.2108$ for the inclination angle, 
polarisation angle, phase, right-ascension, and declination.
The two differ only in the choice of chirp mass and luminosity distance for which we use $30$~$M_{\odot}$ and $500$~Mpc for the \ac{BBH} 
case and $2$~$M_{\odot}$ and $50$~Mpc for the \ac{BNS} case.
The second tuning parameter is $\epsilon$, which sets the threshold loss of SNR.
Choosing $\epsilon$ to be arbitrarily small simply recovers the minimum allowed frequency, i.e. $15$~Hz. We choose $0.1\%$ as a 
conservative estimate, being sufficiently small such that we would not expect any meaningful changes to the inferred parameter estimates.
Finally, we note that this definition has close parallels with the development of varying minimum-frequency bounds in the construction of search template banks \citep{DalCanton:2017ala}.

\begin{figure*}
    \centering
    \includegraphics[width=0.49\textwidth]{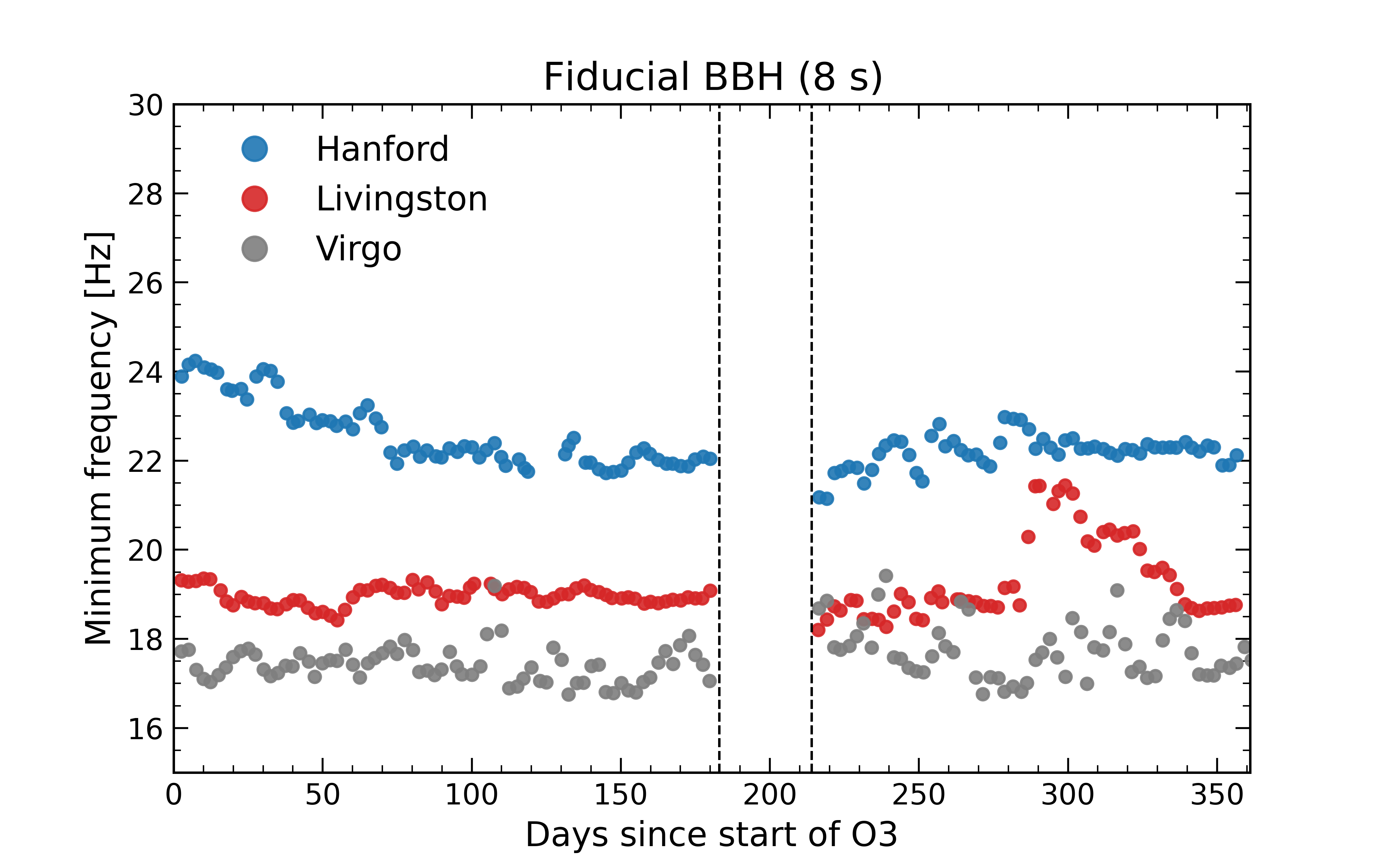}
    \includegraphics[width=0.49\textwidth]{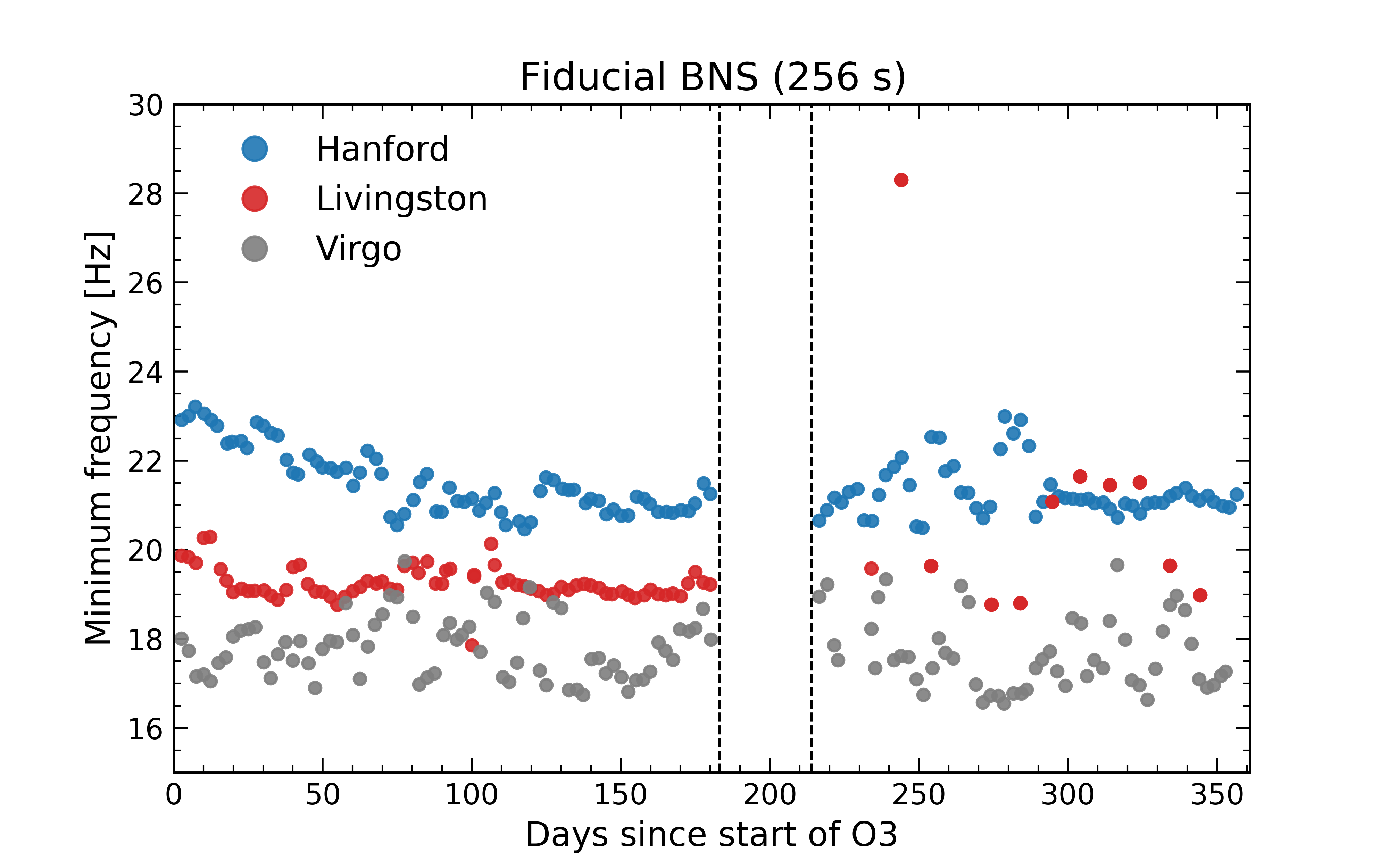}
    \caption{The estimated mean minimum frequency and 90\% interval for LIGO Hanford (H1), LIGO Livingston (L1), and Virgo (V1) over the third observing run for the fiducial \ac{BBH} (left panel) and fiducial \ac{BNS} (right panel). To estimate the mean and interval, we apply a half-overlapping sliding window with a duration of 5 days. Dashed vertical lines mark the beginning and end of the A/B data collection periods.}
    \label{fig:minimum-frequency}
\end{figure*}

Applying these choices, in Fig.~\ref{fig:minimum-frequency}, we plot the estimated value of $\hat{f}_{\rm min}$ in a sliding window
for the three detectors online during the third observing run (O3) of the LIGO-Virgo detector network using data from the Gravitational-Wave Open Science Centre \citep{KAGRA:2023pio}.
We do this first for the fiducial \ac{BBH} (left panel) and then for the fiducial BNS (right panel).
The estimates are time-averaged by the sliding window, smoothing out variations on short timescales 
(e.g. due to the relative sensitivity to the fiducial signal as a function of the rotation of the Earth).
Comparing the detectors, we see that in the \ac{BBH} case, the average for the Hanford instrument is robustly over the $20$~Hz standard, while for Livingston, it averages around $19$~Hz.
This indicates that there may be some minor improvements possible for O3 analyses that use a minimum frequency of $20$~Hz and include 
data from Livingston.
However, we reiterate that our metric is highly conservative (we lose $0.1\%$ of the SNR for a perfectly correlated template), 
and we do not expect the improvements to be significant.
Meanwhile, Virgo has the lowest values of $\hat{f}_{\rm min}$ of the three.
While, in principle, this suggests that using a smaller value of $f_{\rm min}$ could also improve analyses, one should also factor in 
the relative SNR between detectors.
For the fiducial \ac{BBH}, Hanford and Livingston have a median SNR of approximately $30$ while Virgo has a median SNR of $13$; 
for the fiducial BNS, the SNRs are approximately $40$ for Hanford and Livingston, but $20$ for Virgo.
Since the network SNR will be dominated by the more sensitive detectors, for most observations, we expect the standard 
$20$~Hz minimum frequency to be more than sufficient.

We have introduced $\hat{f}_{\rm min}$ as a diagnostic tool to predict when analysts may wish to consider using a minimum frequency.
We note that our choices of tuning parameter mean that $\hat{f}_{\rm min}$ should not be considered an absolute prediction. 
For specific analyses, we suggest performance studies be performed to ensure their choice of $f_{\rm min}$ is robust.
However, we also note that $\hat{f}_{\rm min}$ offers a useful heuristic to understand the relative performance of the detector: 
it, therefore, may be useful in future observing runs as an online monitor or to understand the impact of specific detector improvements.

\section{Discussion}
\label{sec:discussion}
We have simulated gravitational wave signals from compact binary mergers to study the impact of the addition of Virgo and KAGRA 
on the ground-based detector network. We have focused on analyzing the improvements in the sky localization constraints for zero-spin 
binary black holes and neutron star binaries, and on the intrinsic source parameter estimates for high-spin aligned binary black holes.
In Section~\ref{sec:bbh} and~\ref{sec:bns}, we have shown the results of the evolution of the sky-localization area for different 
detector networks when varying the binary neutron star sensitivity range of their least sensitive detector. Our results confirm that 
the addition of the Virgo and \kagra~detectors to the network comprising the two LIGO interferometers would improve the sky localization 
capabilities even when these additional detectors are not at their design sensitivity. Our findings expand on the results from several 
previous studies~\citep{Schutz:2011tw, Fairhurst:2010is, Klimenko:2011hz} that have focused on networks of equally sensitive detectors. 
Looking at the current sensitivities of ground-based interferometers, we find that the inclusion of Virgo into an HLV network is most 
beneficial starting from a Virgo sensitivity range of around 30--50 Mpc, corresponding to $\sim$1/5 of the sensitivity range of the LIGO 
detectors, that we fixed at 180~Mpc. These results, which depend on the specific position of the source relative to the Virgo detector and its 
inclination, are consistent with the relative sensitivity values of $\simeq 1/5-1/6$ that we computed in Section~\ref{sec:Method}.
We find that our 
results also agree with the analysis of the binary black hole merger signal GW170814~\citep{LIGOScientific:2017ycc}. For this first clear 
three-detector observation, the 90\% sky-localization area improves from 1160 to 60 deg$^2$ when adding the Virgo detector data, whose 
sensitivity range at the time was roughly 1/3 of the LIGO detector's sensitivity. Although the source of GW170814 was much nearer 
than our simulated one, this is comparable to the improvement we see in Figure~\ref{fig:HLV_bbh} for 60 Mpc, i.e., 10 deg$^2$ 
compared to the 800 deg$^2$ of the HL network.

A more general heuristic consideration is to look at the cumulative distribution 
of the antenna pattern function of the two individual LIGO detectors. When these are marginalized over all four angular parameters, one 
can notice that Virgo is almost exactly orthogonal to the LIGO detectors, as seen in Figure~\ref{fig:antenna_power_pattern}. 
So, it would have the same sensitivity as LIGO, or better, for 
about half of all the sources even if its sensitivity range was a factor of two smaller.  Since we can argue that it would be enough 
to observe sources at about 1/3 of the LIGO signal-to-noise ratio in Virgo to still have meaningful parameter estimation, then Virgo 
would need to have at least a sensitivity a sixth that of LIGO, or $\simeq$30~Mpc average for about half of the sources.
Comparing the field of view values of current optical telescopes, presented in Section~\ref{sec:Intro}, with the results of our simulations, we notice that the Virgo detector would allow for an easier 
follow-up from 30 Mpc and 90 Mpc onwards, for the HLV and HV detector networks respectively, for a binary neutron star signal. In 
the non-spinning binary black hole case, the sky localization area is reduced to $\sim$10 deg$^2$ when Virgo is added to the HLV 
network with a sensitivity above 45 Mpc. For all the other two- and three-detector networks, despite the significant improvements in the 
localization constraints, these do never fall below the 10 deg$^2$ threshold. It is still noteworthy that the addition of \kagra, 
with a sensitivity of 25 Mpc, into the HLK network reduces the inferred sky area from 200 deg$^2$ to 35 deg$^2$ for a binary neutron 
star source. From our four-detector network results, shown in Figure~\ref{fig:HLVK_bbh} and~\ref{fig:HLVK_ns}, we see that when Virgo 
has a sensitivity of 30 Mpc, 1/6 of LIGO's, the addition of the \kagra~detector is still useful to constraint the sky-area of events 
maximizing \kagra's antenna pattern function. Specifically, for the binary neutron star case, the sky area can be constrained to under 
$\sim$35 deg$^2$ when \kagra~reaches a sensitivity greater than 20 Mpc, corresponding to $\sim$1/9 of LIGO's. The mentioned results 
are obtained for a specific case, i.e., we consider a maximized Virgo antenna pattern function and a fixed distance, but our choice 
of parameters is a good approximation of the average values for the currently detectable population of sources~\citep{Kagra:2021duu}. Furthermore, 
our focus on the ratios of relative detector sensitivities in a network easily allows for a translation of our results to future 
detector networks. 

In our analysis, we also focused on two detector networks considering the variable duty factor the two LIGO detectors had in the 
previous observation runs, e.g., 65.3\% and 61.8\% during O2 for Hanford and Livingston, respectively~\citep{Kagra:2013rdx}. We found that even in the two detector cases, the addition of Virgo and KAGRA can have an important impact on the performance of the network. With Virgo, at a sensitivity of 100~Mpc or higher, we can expect the sky localization to be constrained to values smaller than the field view of current optical telescopes. This would highly increase our chances of detecting a multi-messenger event. 

\section{Conclusion}
\label{sec:conclusion}
We have run parameter estimation with \Bilby on simulated gravitational wave signals from zero-spin black hole and neutron star binaries looking at the sky localization capability of different detector network configurations. We have focused on the improvements related to the addition of the Virgo and \kagra~detectors. We kept the sensitivity of the LIGO detectors fixed at 180 Mpc and varied the sensitivity of the additional detector. We found that the Virgo detector allows for the most significant improvements in the sky localization when it reaches a sensitivity that is approximately 1/6 that of the LIGO detectors. We expect a similar result for \kagra, but we restrict our study to values up to 1/7 that of LIGO. Furthermore, we found that in the four-detector case, the addition of \kagra~would be beneficial at sensitivities above 1/10 that of the LIGO detectors, if Virgo is at 1/6 of the LIGO sensitivity.\\ We also simulated gravitational waves from high-spin aligned binary black holes and analyzed the improvements in the inference of the intrinsic source parameters with the sensitivity of the Virgo detector. We found the most significant improvements in the accuracy at sensitivities of the Virgo detector above half that of the LIGO detectors. \\ Finally, we have developed a tool to determine the optimal minimum frequency, defined as the lowest frequency for which the signal-to-noise ratio does increase by more than 0.1\% with respect to higher cut-off frequency values. Testing this on O3 data by adding simulated gravitational wave signals, we found that 20 Hz is a good choice for the lower frequency cut-off in the parameter estimation analysis.

Finally, we have applied a naive scaling of the spectral density, but we have a good understanding of the noise budget of the detectors as a combination of power laws. In the future, we could use this knowledge to develop an interface to track the real-time changes in the PSD with the improvements in the detector. The parameters of interest would be the fitting constants for each power law in the PSD. This could allow us to overcome the imperfect scaling used in this work and would give us a deeper understanding of the connection between detector improvements and noise curves. 

\begin{acknowledgments}
We would like to thank Sathyaprakash Bangalore for input into the analytic predictions for network sensitivity improvements and Carl-Johan Haster for feedback which helped improve the clarity of the exposition.
The authors are grateful for computational resources provided by the LIGO Laboratory and supported by National Science Foundation Grants PHY-0757058 and PHY-0823459.
The material in Section~/\ref{sec:min_freq} is based upon work supported by NSF's LIGO Laboratory which is a major facility fully funded by the National Science Foundation.
This research has made use of data or software obtained from the Gravitational Wave Open Science Center (gwosc.org), a service of the LIGO Scientific Collaboration, the Virgo Collaboration, and KAGRA. This material is based upon work supported by NSF's LIGO Laboratory which is a major facility fully funded by the National Science Foundation, as well as the Science and Technology Facilities Council (STFC) of the United Kingdom, the Max-Planck-Society (MPS), and the State of Niedersachsen/Germany for support of the construction of Advanced LIGO and construction and operation of the GEO600 detector. Additional support for Advanced LIGO was provided by the Australian Research Council. Virgo is funded, through the European Gravitational Observatory (EGO), by the French Centre National de Recherche Scientifique (CNRS), the Italian Istituto Nazionale di Fisica Nucleare (INFN) and the Dutch Nikhef, with contributions by institutions from Belgium, Germany, Greece, Hungary, Ireland, Japan, Monaco, Poland, Portugal, Spain. KAGRA is supported by Ministry of Education, Culture, Sports, Science and Technology (MEXT), Japan Society for the Promotion of Science (JSPS) in Japan; National Research Foundation (NRF) and Ministry of Science and ICT (MSIT) in Korea; Academia Sinica (AS) and National Science and Technology Council (NSTC) in Taiwan.

\end{acknowledgments}

\newpage

\bibliography{biblio.bib}

\appendix

\onecolumngrid
\section{Tables of simulation parameters and prior distributions}
\label{sec: app_tables}

\begin{table*}[htp!]
\caption{\label{tab:simulation} Parameter values for simulated gravitational wave signals corresponding to non-spinning \ac{BBH}, \ac{BNS}, and high-spin \ac{BBH} mergers. Values are given for the chirp mass ($\mathcal{M}$), mass ratio (q), luminosity distance (D$_L$), dimensionless spins (a$_1$ and a$_2$), tilt angles (tilt$_1$ and tilt$_2$), phase angles ($\phi_{12}$ and $\phi_{jl}$), angular parameter ($\theta_{jn}$), polarization angle ($\psi$), and overall phase. These simulated values serve as the ground truth for the simulated signals.}
\begin{ruledtabular}
\begin{tabular}{cccc}
 Parameter&BBH &BNS& High-spin BBH
 \\   \hline 
 $\mathcal{M}$& 50 &1.198 &27.93 \\
 q& 1 & 1 &0.40\\
 D$_L$& 2000 & 100 & 2000
 \\
 a$_1$&0& 0 &0.91\\
 a$_2$&0 &0&0.94 \\
 tilt$_1$&0 &0& 0 \\
 tilt$_2$&0 &0& 0 \\
 $\phi _{12}$&0 &0&0 \\
 $\phi _{jl}$&0 &0&0 \\
 $\theta _{jn}$&0.4 &0.4&0.4 \\
 $\psi$&2.66 &2.66& 2.66 \\
  $\phi$&1.3 &1.3&1.3 \\
\end{tabular}
\end{ruledtabular}
\end{table*}

\begin{table*}[htp!]
\caption{\label{tab:prior}Prior distributions employed in the parameter estimation of simulated gravitational wave signals arising from high-spin and non-spinning \ac{BBH} and \ac{BNS} mergers. The symbol $\mathcal{U}$ denotes a uniform prior, and $\lambda^2$ represents a power law prior within the specified ranges. The table details the priors for chirp mass ($\mathcal{M}$), mass ratio (q), individual masses ($M_1$ and $M_2$), dimensionless spins ($\chi_1$ and $\chi_2$), tilt angles (tilt$_1$ and tilt$_2$), declination (DEC), luminosity distance (D$_L$), right ascension (RA), and various angle parameters..}
\begin{ruledtabular}
\begin{tabular}{cccc}
 Parameter&BBH &BNS& High-spin BBH
 \\   \hline \vspace{-0.2cm} \\
 $\mathcal{M}$& $\mathcal{U}$ [25,60] &$\mathcal{U}$ [1.19799,1.19801] &$\mathcal{U}$ [15,60] \\
 q& $\mathcal{U}$ [0.125,1]& $\mathcal{U}$ [0.125,1] &$\mathcal{U}$ [0.125,1]\\
 M$_1$& [1,100]& [1,100] &[5,100]\\
 M$_2$& [1,100]& [1,100] &[5,100]\\
 $\chi_1$& $\mathcal{U}$ [0,0.99]& $\mathcal{U}$ [0,0.05] & -\\
 $\chi_2$& $\mathcal{U}$ [0,0.99]& $\mathcal{U}$ [0,0.05] &-\\
 $\mathrm{a}_1$& -& - &$\mathcal{U}$ [0,0.99]\\
 $\mathrm{a}_2$& -& - &$\mathcal{U}$ [0,0.99]\\
  tilt$_1$& -& - &$\sin$\\ 
  tilt$_2$& -& - &$\sin$\\
  $\phi_{12}$& -& - &$\mathcal{U}$ [0,2$\pi$]\\
  $\phi_{jl}$& -& - &$\mathcal{U}$ [0,2$\pi$]\\
 D$_L$& $\lambda^2$ [10,10000] & $\lambda^2$ [10,500] & $\mathcal{U}$ [100,5000] \\
 DEC& $\cos$& $\cos$ &$\cos$\\
 RA&$\mathcal{U}$ [0,2$\pi$] & $\mathcal{U}$ [0,2$\pi$] &$\mathcal{U}$ [0,2$\pi$]\\
 $\theta_{jn}$&$\sin$ &$\sin$& $\sin$ \\
 $\psi$& $\mathcal{U}$ [0,$\pi$]&  $\mathcal{U}$ [0,$\pi$] &  $\mathcal{U}$ [0,$\pi$]\\
 $\phi$& $\mathcal{U}$ [0,2$\pi$]& $\mathcal{U}$ [0,2$\pi$] & $\mathcal{U}$ [0,2$\pi$]\\
\end{tabular}
\end{ruledtabular}
\end{table*}

\begin{table*}[htp!]
\caption{\label{tab:dinesty} Setting employed for the \dynesty sampler of \Bilby for the parameter estimation of the simulated gravitational wave signals.}
\begin{ruledtabular}
\begin{tabular}{ccccc}
\dynesty & nlive & npool& sample & naccept
 \\   \hline 
zero-spin BBH/BNS & 1000 & 30 &acceptance-walk& 60 \\
high-spin BBH & 2000 & 30 &acceptance-walk& 60 \\
\end{tabular}
\end{ruledtabular}
\end{table*}

\newpage
\onecolumngrid
\section{Figures and table for the best network sky localization}
\label{sec: best network}

\begin{table*}[htp!]
\centering
\begin{tabular}{|c|c|c|c|c|c|c|} \hline 

                & RA (rad)  &  DEC (rad)  & P$_{K}$ & P$_{V}$ & P$_{H}$ & P$_{L}$\\ \hline  
HK~P$_{\Pbest}$      & 5.776        &   -0.952 & 0.62 & 0.17 & 0.72 & 0.32\\ \hline    
HV~P$_{\Pbest}$       & 5.311        &   0.917 & 0.10 & 0.88 & 0.35 & 0.34\\ \hline  
HLK P$_{\Pbest}$       & 3.678        &   0.638  & 0.21 & 0.1 & 0.95 & 0.91\\ \hline  
HLV P$_{\Pbest}$       & 4.008        &   0.690  & 0.10 & 0.23 & 0.87 & 0.98\\ \hline
HLVK P$_{\Pbest}$       & 3.873        &   0.529  & 0.20 & 0.20 & 0.85 & 0.98\\ \hline
\end{tabular}
\caption{\RaggedRight \justifying Table of the right ascension (RA), declination (DEC) and the values of the single
detector's antenna pattern amplitude for the source 
localization maximizing the combined antenna power pattern function, P$_{\Pbest}$, 
for each of the network configurations. The GPS time is fixed at 1379969683.0. }
\label{tab:localization_best}
\end{table*}

\begin{figure*}[htp!]
\centering
\subfloat[\label{fig:HLK_bbh_best}]{%
     \includegraphics[width=0.46\textwidth]{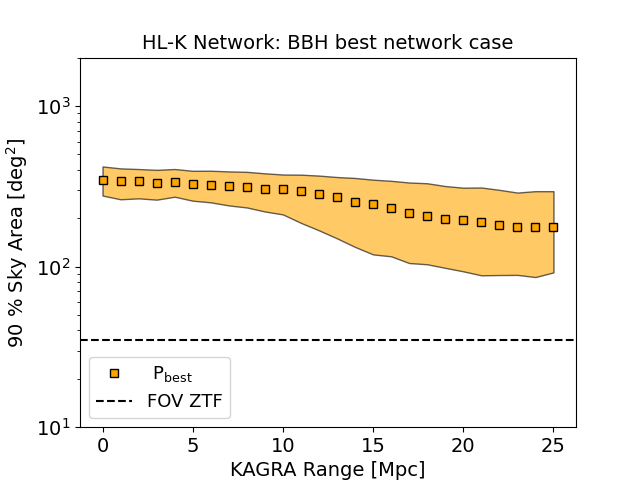}}%
\hfil
\subfloat[\label{fig:LK_bbh_best}]{%
   \includegraphics[width=0.46\textwidth]{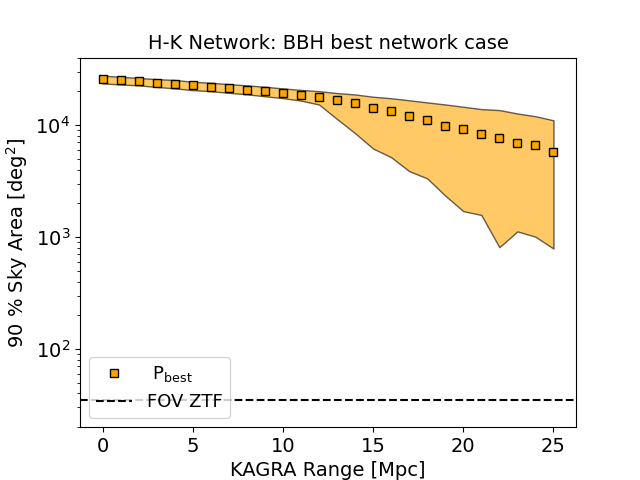}}%
\vskip\floatsep

\subfloat[\label{fig:HLV_bbh_best}]{
    \includegraphics[clip,width=0.46\textwidth]{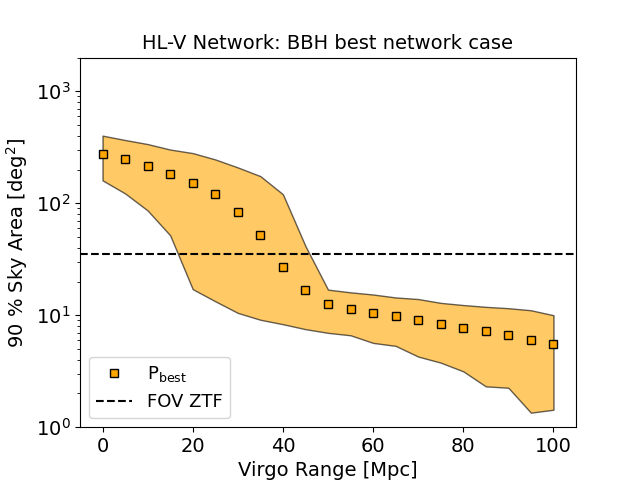}}%
\subfloat[\label{fig:LV_bbh_best}]{
    \includegraphics[clip,width=0.46\textwidth]{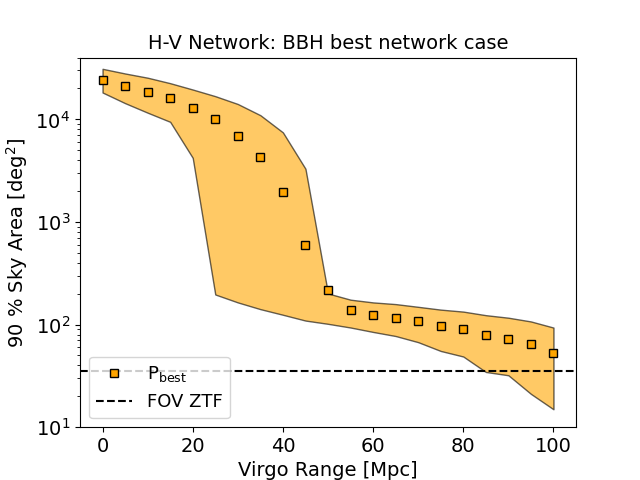}}%
\vskip\floatsep    
\subfloat[\label{fig:HLVK_bbh_best}]{
    \includegraphics[clip,width=0.46\textwidth]{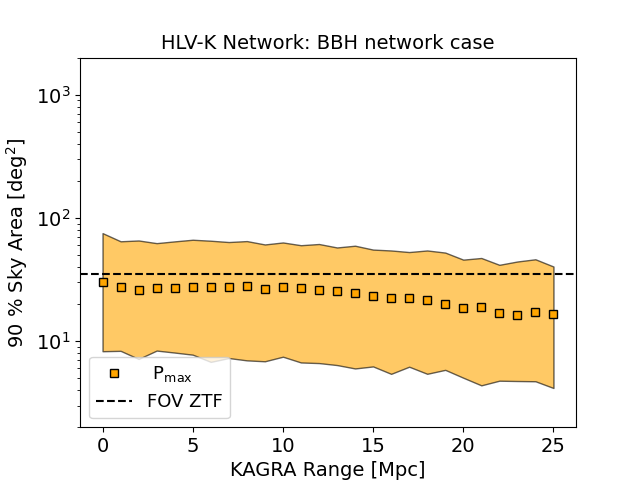}}%
\sbox0{%
\begin{minipage}[b]{0.46\textwidth}%
\mycaption{Zero-Spin Binary Black Hole Simulations}{ \RaggedRight \justifying The mean and the 90 \% probability contour of the sky area for varying detector range sensitivity values for a gravitational wave signal emitted 
by a binary black hole coalescence from the P$_{\Pbest}$ source location for 
each detector network, orange points and area, respectively. The orange squares 
represent the mean value, 
and the areas stretch from the 5\% to the 95\% quantiles for each sensitivity range bin. The black dotted line corresponds to the field of view of the Zwicky Transient Facility. We obtained the 0 Mpc 
points from simulations excluding the least sensitive detector. We fixed the range sensitivity of the two LIGO 
detectors at 180 Mpc. \textbf{(a)}: Varying the \kagra~detector and employing the LHK network. \textbf{(b)}: 
Varying the \kagra~detector and employing the HLK network. \textbf{(c)}: Varying the Virgo detector and employing 
the LHV network. \textbf{(d)}: Varying the Virgo detector and employing the HLV network. \textbf{(e)}: Varying the 
\kagra~detector and employing the LHVK network. Virgo's sensitivity is fixed at 30 Mpc.  }
\label{fig:BBH_best}
\end{minipage}
}%
\hfil
\usebox0
\end{figure*} 

\begin{figure*}[htp!]
\centering
\subfloat[\label{fig:HLK_ns_best}]{%
     \includegraphics[width=0.46\textwidth]{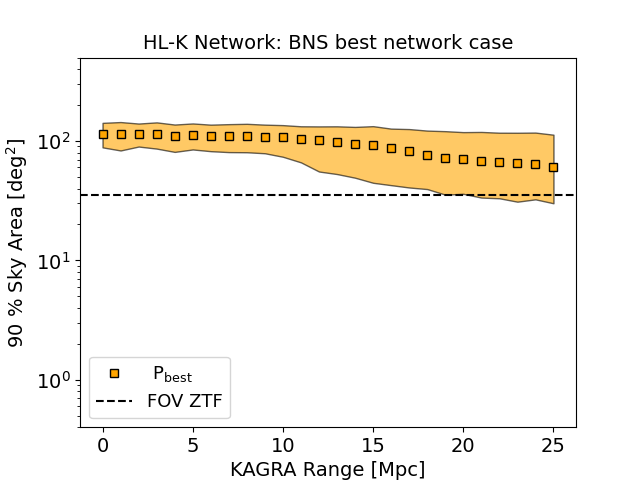}}%
\hfil
\subfloat[\label{fig:LK_ns_best}]{%
   \includegraphics[width=0.46\textwidth]{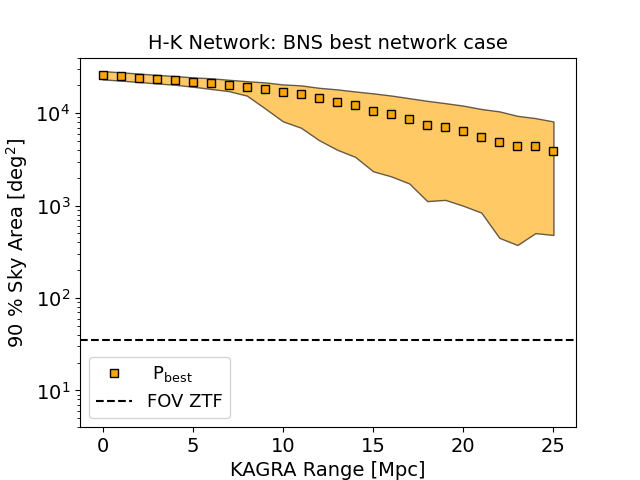}}%
\vskip\floatsep

\subfloat[\label{fig:HLV_ns_best}]{
    \includegraphics[clip,width=0.46\textwidth]{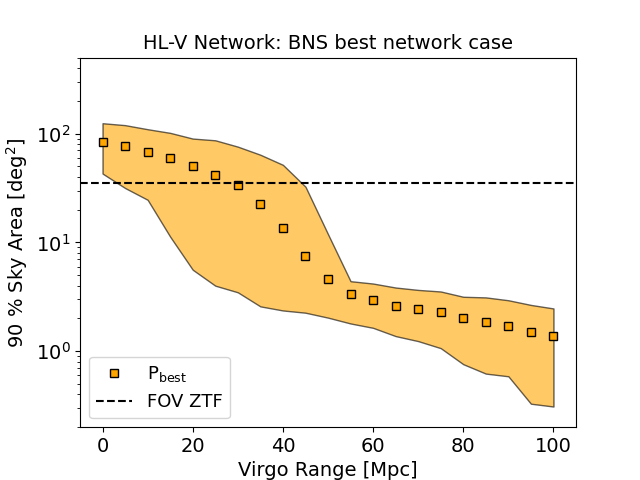}}%
\subfloat[\label{fig:LV_ns_best}]{%
    \includegraphics[clip,width=0.46\textwidth]{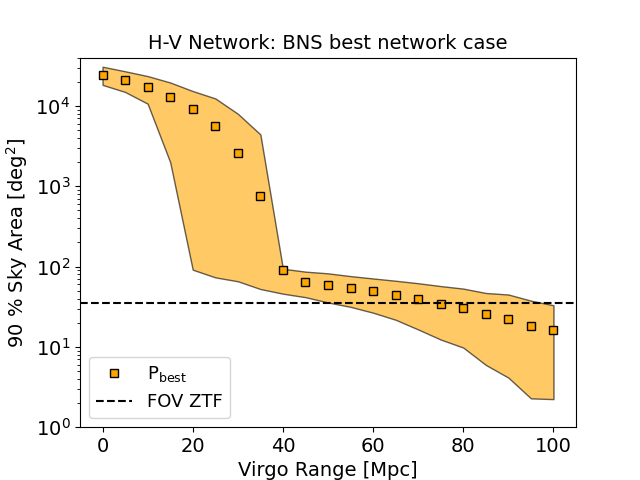}}%
\vskip\floatsep    
\subfloat[\label{fig:HLVK_ns_best}]{
    \includegraphics[clip,width=0.46\textwidth]{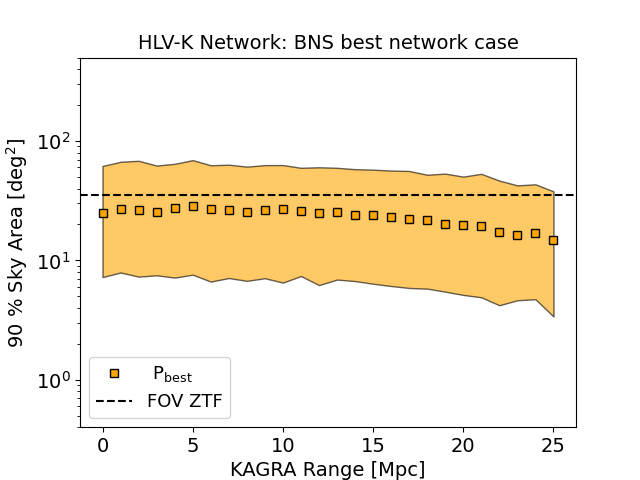}}%
\sbox0{%
\begin{minipage}[b]{0.46\textwidth}%
\mycaption{Binary Neutron Star Simulations}{ \RaggedRight \justifying The mean and the 90\% probability contour 
of the sky area for varying detector range sensitivity values for a gravitational wave signal emitted by a 
binary neutron star coalescence from the P$_{\Pbest}$ source location for 
each detector network, orange dots and areas, respectively. The squares represent the mean value, 
and the areas stretch from the 5\% to the 95\% quantiles for each sensitivity range bin. The black dotted line corresponds to the field of view of the Zwicky Transient Facility. We obtained the 0~Mpc 
points from simulations excluding the least sensitive detector. We fixed the range sensitivity of the two LIGO 
detectors at 180~Mpc. \textbf{(a)}: Varying the \kagra~detector and employing the LHK network. \textbf{(b)}: 
Varying the \kagra~detector and employing the HLK network. \textbf{(c)}: Varying the Virgo detector and employing 
the LHV network. \textbf{(d)}: Varying the Virgo detector and employing the HLV network. \textbf{(e)}: Varying 
the \kagra~detector and employing the LHVK network. Virgo's sensitivity is fixed at 30~Mpc.}
\label{fig:BNS_best}
\end{minipage}
}%
\hfil
\usebox0
\end{figure*} 
\end{document}